\documentclass[acmsmall,screen]{acmart}

\acmJournal{PACMPL}
\acmVolume{OOPSLA}
\acmNumber{1}
\acmArticle{1}
\acmMonth{1}

\setcopyright{rightsretained}
\acmJournal{PACMPL}
\acmYear{2021} \acmVolume{5} \acmNumber{OOPSLA} \acmArticle{131} \acmMonth{10} \acmPrice{}\acmDOI{10.1145/3485508}



\usepackage{xcolor}
\usepackage{graphicx}
\usepackage{xspace}
\usepackage{listings}
\usepackage{mathtools}
\usepackage{pgfplots}
\usepackage[pdf]{graphviz}
\usepackage{color, colortbl}
\usepackage{subfigure}
\usepackage{tabularx}
\usepackage{wrapfig}

\newcolumntype{L}[1]{>{\raggedright\let\newline\\\arraybackslash\hspace{0pt}}m{#1}}

\lstdefinelanguage{LNT}
{alsoletter={\[,\]},
morekeywords={while,for,by,process,any,is,var,in,loop,select,end,any,\[\]},
sensitive=false,
morecomment=[l]{--},
morecomment=[s]{(*}{*)},
morestring=[b]",
numbers=left,
basicstyle=\ttfamily\footnotesize,
}

\lstdefinelanguage{MCL}
{morekeywords={for,from,to,do,or,end,if,then,true,is_in,where,is_empty,not,<,>,@},
basicstyle=\ttfamily\footnotesize,
}

\lstdefinelanguage{DSL}
{morekeywords={AXB},
morecomment=[l]{//},
basicstyle=\ttfamily\footnotesize,
}

\definecolor{Gray}{gray}{0.9}

\newcommand{\includeappendix}{defined}

\usepackage[normalem]{ulem}

\newcommand{\edit}[1]{#1}

\newcommand{\numtests}{483}
\newcommand{\numvendors}{5}
\newcommand{\numgpus}{8}

\newcommand{\rtx}{\textsc{rtx4k}\xspace}
\newcommand{\gfm}{\textsc{gf940m}\xspace}
\newcommand{\anew}{\textsc{a14}\xspace}
\newcommand{\aold}{\textsc{a12}\xspace}
\newcommand{\mali}{\textsc{g77}\xspace}
\newcommand{\adreno}{\textsc{a620}\xspace}
\newcommand{\tegra}{\textsc{tx1}\xspace}
\newcommand{\intel}{\textsc{hd620}\xspace}

\newcommand{\atomicxb}{\texttt{ATOMIC\_EXCH\_BRANCH}\xspace}
\newcommand{\axb}{\texttt{AXB}\xspace}

\def\Snospace~{\textsection{}}

\begin{document}

\title{Specifying and Testing GPU Workgroup Progress Models}

\author[T. Sorensen]{Tyler Sorensen}
\affiliation{%
  \institution{UC Santa Cruz}
  \country{USA}
  }
\email{tyler.sorensen@ucsc.edu}
  
\author[L. F. Salvador]{Lucas F. Salvador}
\affiliation{%
  \institution{Princeton University}
  \country{USA}
  }
\email{ls24@alumni.princeton.edu}

\author[H. Raval]{Harmit Raval}
\affiliation{%
  \institution{Princeton University}
  \country{USA}
  }
\email{hraval@alumni.princeton.edu}
      
\author[H. Evrard]{Hugues Evrard}
\affiliation{%
  \institution{Google}
  \country{UK}
  }
\email{hevrard@google.com}
  
\author[J. Wickerson]{John Wickerson}
\affiliation{%
  \institution{Imperial College London}
  \country{UK}
  }
\email{j.wickerson@imperial.ac.uk}
  
\author[M. Martonosi]{Margaret Martonosi}
\affiliation{%
  \institution{Princeton University}
  \country{USA}
  }
\email{mrm@princeton.edu}  
  
\author[A. F. Donaldson]{Alastair F. Donaldson}
\affiliation{%
  \institution{Imperial College London}
  \country{UK}
  }
\email{alastair.donaldson@imperial.ac.uk} 

\date{}

\thispagestyle{empty}

\begin{abstract}

As GPU availability has increased and programming support has matured, a wider variety of applications are being ported to these platforms. Many parallel applications contain fine-grained synchronization idioms; as such, their correct execution depends on a degree of relative forward progress between threads (or thread groups). Unfortunately, many GPU programming specifications (e.g.\ Vulkan and Metal) say almost nothing about relative forward progress guarantees between workgroups.
Although prior work has proposed a spectrum of plausible progress models for GPUs, cross-vendor specifications have yet to commit to any model.

This work is a collection of tools and experimental data to aid specification designers when considering forward progress guarantees in programming frameworks. As a foundation, we formalize a small parallel programming language that captures the essence of fine-grained synchronization. We then provide a means of formally specifying a progress model, and develop a termination oracle that decides whether a given program is guaranteed to eventually terminate with respect to a given progress model. Next, we formalize a set of constraints that describe concurrent programs that require forward progress to terminate. This allows us to synthesize a large set of \numtests{} progress litmus tests. Combined with the termination oracle, we can determine the expected status of each litmus test -- i.e.\ whether it is guaranteed to eventually terminate -- under various progress models.  We present a large experimental campaign running the litmus tests across \numgpus{} GPUs from \numvendors{} different vendors. Our results highlight that GPUs have significantly different termination behaviors under our test suite. Most notably, we find that Apple and ARM GPUs do not support the \emph{linear occupancy-bound} model, an intuitive progress model defined by prior work that has been hypothesized to describe the workgroup schedulers of existing GPUs. 
\end{abstract}

\maketitle

\section{Introduction \label{sec:intro}}
Since its inception nearly a decade ago in frameworks such as CUDA and OpenCL, \edit{general-purpose computing on graphics processing units (GPGPU computing) has seen two trends:} more and more applications are offloading computations to GPU devices, and the diversity and availability of GPU architectures has increased dramatically. Almost all major chip designers, e.g.\ Nvidia, Intel, AMD, ARM, Qualcomm, and Apple, provide their own GPU, and nearly every mainstream computing device contains a programmable GPU.

While some GPU vendors provide programming frameworks specific to their devices (notably Nvidia with CUDA and Apple with Metal), the diversity of the modern computing landscape motivates cross-vendor GPU programming frameworks. Indeed, OpenCL has recently launched a overhauled version 3.0 with many modern features, including support for C++17 and integration with Clang~\cite{opencl-3-release}. And at a lower level, Vulkan~\cite{vulkan} was released in 2016, is supported by most vendors, and has had considerable uptake by users.

\paragraph{Independent forward progress on GPUs} Given implementation differences across GPUs, cross-vendor GPU frameworks differ from those of classic CPU programming.
In particular, \emph{independent forward progress} (henceforth referred to as \emph{progress}) between threads of execution is not always guaranteed on a GPU.  This means that it is not safe---in general---to write a GPU program (often called a \textit{kernel}) where one thread relies on another thread making progress.
Full progress guarantees are difficult to provide for two main reasons.
First, a GPU computation is split into sub-tasks, each of which is assigned to a subset of threads called a \emph{workgroup}.  If there are too many workgroups, they might not all execute concurrently.  The GPU scheduler might delay execution of some workgroups until others have finished. If a thread from an early workgroup spins, waiting for a thread from a later, delayed, workgroup, this will lead to starvation and execution will not terminate. 
Second, the threads of a workgroup are usually organized into \emph{subgroups} where each thread in a subgroup is really a lane in a stream of SIMD vector instructions.  The threads in a subgroup share a program counter and thus do not progress independently.

\paragraph{A case for supporting progress} 
Given this, many GPGPU framework specifications do not commit to any progress guarantees. Even the Nvidia-specific CUDA framework has only recently provided progress guarantees, requiring specialized hardware support~\cite[p. 26]{VoltaWhitepaper}. The lack of cross-vendor progress guarantees severely limits the kinds of concurrent programs that can be safely offloaded onto GPUs, so that even fundamental synchronization idioms, such as mutexes and execution barriers, are not supported in principle and have been shown to lead to non-terminating behavior in practice~\cite{GPUBarrierOOPSLA2016,MIMDSyncOnSIMD}.
We argue that GPU programming specifications \emph{should} provide progress guarantees in some form, for several reasons:

\begin{itemize}

    \item \textbf{Current devices empirically provide progress.} 
    Prior work has empirically shown that today's GPUs appear to guarantee a limited form of progress between workgroups~\cite{GPUBarrierOOPSLA2016,Sorensen-Pai-Donaldson-19}. %These works have run applications containing specialized synchronization constructs for many hours without issue. 
    The putative progress model, \textit{Occupancy-Bound Execution} (OBE), states that any workgroup that starts execution will be fairly scheduled until it finishes execution.
  
    \item \textbf{Developers already rely on empirical guarantees.} Examples of developers relying on unofficial progress models for GPUs goes back to 2010, when a limited execution barrier was proposed~\cite{XF10}. Many examples followed, e.g.\ mutexes and workstealing, that rely on some progress model, typically OBE~\cite{TPO10,ct08,MYB16,dont-forget-about-sync}. These folklore assumptions have had a decade to take root, but cross-vendor GPU programming frameworks still do not provide a means of querying whether such assumptions hold for a given device.

    \item \textbf{Progress guarantees can enable significant performance.} These GPU folklore progress models are required for several high-performance GPU programs. For example, a graph application optimization requiring OBE progress properties has been shown to drastically improve performance across many devices~\cite{Sorensen-Pai-Donaldson-19}. Additionally, the high-performance prefix sum implementation~\cite{BlellochScan, Merry2015APC} in the CUB CUDA library~\cite{cub} also requires inter-workgroup progress support; it was recently ported to Vulkan and shown to have high performance across a range of devices~\cite{raph-blog}.

    \item \textbf{Progress guarantees are inconsistent across frameworks.}
    Nvidia's CUDA framework recently introduced a progress model that legitimises some of these folklore assumptions. There is a risk that developers familiar with CUDA may wrongly assume that its guarantees apply more broadly, especially when using compilation flows that translate from CUDA into other frameworks for the explicit purpose of porting CUDA code to other devices~\cite{cu2cl,cuda-to-opencl}. Bringing non-trivial notions of progress guarantees to cross-vendor frameworks would help to close this gap and avoid such problems.

\end{itemize}

Despite being aware of these considerations, designers of GPGPU frameworks have reasons \emph{not} to impose blanket progress guarantees: doing so might be costly or infeasible on some architectures, or may have unanticipated consequences for future architectures.
Yet, progress guarantees do not have to be all-or-nothing. Previous research on \emph{semi-fair} schedulers~\cite{Sorensen-Evrard-Donaldson-18} defines a  hierarchy of progress models that offer a range of progress guarantees.
Furthermore, progress guarantees could be incorporated as queryable features of a programming framework, allowing devices to advertise the particular forward progress models that they support.

\subsection*{Contributions} To make informed decisions about progress models, framework designers must be able to: (1)~\textit{describe} forward progress models, (2)~\textit{ask questions} about the subtle concurrent interactions that a given model allows, and (3)~\textit{test} whether an implementation satisfies the chosen forward progress model. Our contributions, which aim to support designers in these regards, are as follows:

\paragraph{A simple programming language for progress reasoning (\autoref{sec:programming-model})}
We define a small concurrent programming language
designed specifically for expressing progress model litmus tests: small unit tests that are guaranteed to terminate under a strongly fair progress model, but that have the potential for non-termination under more restrictive models.
Featuring just \textit{one} type of instruction---an atomic instruction that encodes a read, an optional write, and an optional branch---the language is simple enough to allow reasoning and test synthesis using off-the-shelf tools, yet is rich enough to express concurrent idioms such as producer-consumer, mutexes, and barriers; the kinds of examples that inform discussion of progress models.

\paragraph{Executable semantics for progress models (\autoref{sec:cadp})}
We show that a progress model can be rigorously described in a process-algebraic formalism over the programming language of \autoref{sec:programming-model}.
For the user, i.e.\ the specification designer, this involves specifying the set of threads that are guaranteed eventual execution at various points in the program execution.

We provide an implementation in the CADP model checker~\cite{CADP2011} that can be used to answer questions about whether concurrent programs written in our language are guaranteed to terminate under various progress models. We provide a library of 6 progress models of interest, including all that we were able to find in prior work.
Of these 6 progress models, 5 are influenced by the fairness being weak or strong, which brings the total to 11 distinct progress models.

\paragraph{Synthesising progress litmus tests (\autoref{sec:alloy})} 

We formalize a set of constraints that progress litmus tests must satisfy; i.e.\ constraints ensuring that a program will terminate under a fair progress model but is not guaranteed to terminate in the absence of progress guarantees.
By expressing these constraints in the Alloy modeling language~\cite{alloy} and feeding them to the Alloy Analyzer we are able to synthesize a large set of progress litmus tests.
Combined with the executable semantics, this facilitates exploration of progress models, e.g.\ discovering interesting and unexpected programs that are not guaranteed to terminate under a particular model, or that show subtle differences between two similar models.
Our approach allows us to synthesize a set of \emph{conformance tests} for a given progress model, which a correct implementation of the model should be expected to pass.
We have used this to automatically synthesise a suite of \numtests{} progress litmus tests. Inside this test suite we have identified tests that correspond to common synchronization constructs such producer-consumer and mutex idioms. Our test suite is able to distinguish 10 out of our 11 example progress models; i.e.\ the set of correct/incorrect progress litmus tests are different for each model.

\paragraph{Evaluating the progress models of GPUs in the field (\autoref{sec:testing})}
We have developed three back-ends that transform a high-level progress litmus test generated by the Alloy-based approach of \autoref{sec:alloy} into an executable GPU program, supporting the Vulkan, CUDA and Metal programming frameworks.
We present a large experimental campaign running our suite of \numtests{} progress litmus tests against \numgpus{} GPU devices from \numvendors{} vendors, including discrete, integrated and mobile GPUs. 
Similar to prior works~\cite{Sorensen-Evrard-Donaldson-18,Dutu-Sinclair-Beckmann-Wood-Chow-20}, our testing efforts focus on inter-workgroup scheduling.
Our results confirm a hypothesis from prior work~\cite{Sorensen-Evrard-Donaldson-18} that all current GPUs support the OBE progress model, as well as an incomparable progress model associated with the HSA GPU programming framework~\cite{HSABook}. Interestingly our results \emph{refute} a stronger hypothesis from this prior work: that current GPUs support a progress model known as LOBE, a natural combination of the OBE and HSA models. On GPUs from Apple and from ARM, we observe non-termination behavior for tests that are guaranteed to pass under the LOBE model.

\begin{figure}[t]
    \centering
    \includegraphics[width=\textwidth]{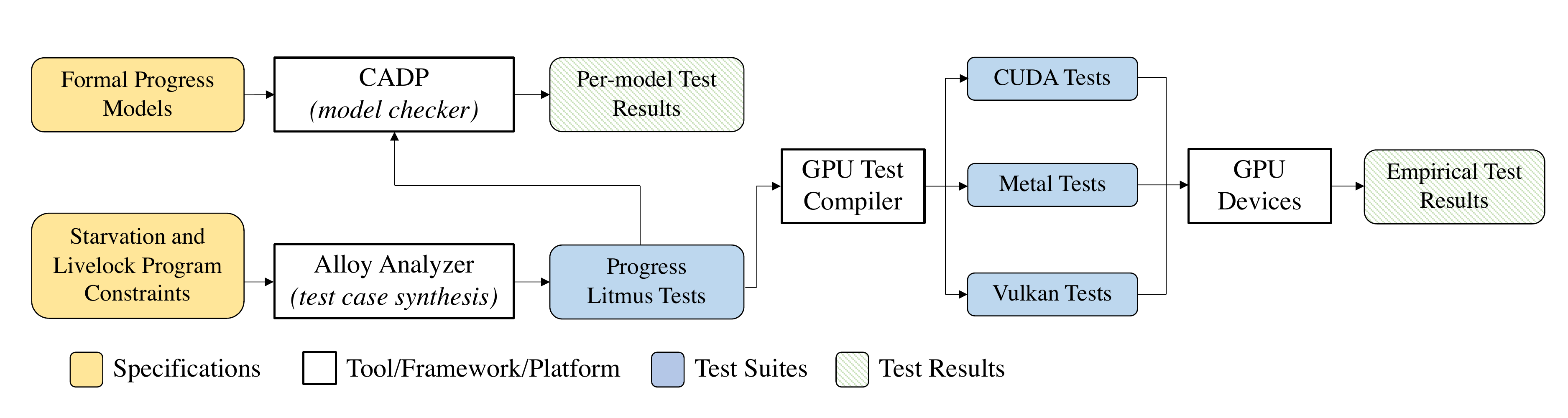}
    \caption{\edit{A flowchart detailing our approach. We present two types of specifications: one to describe our progress models, and one that defines constraints over programs that contain potential non-termination due to starvation or livelock. The program constraint specification can be fed to the Alloy Analyzer for progress litmus test synthesis. The progress litmus tests can take two routes: they can be executed in the CADP model checker under any of our formal progress model specifications, providing formal per-model test outcomes. They can also be input into our GPU test compiler, which outputs executable test cases for three common GPGPU frameworks. The formal test results can be cross-referenced with the empirical test to explore which (if any) formal progress model(s) describe the observed behavior across different GPUs. }}
    \label{fig:flowchart}
\end{figure}

\medskip
By providing a practical toolchain (\edit{detailed in \autoref{fig:flowchart}}) to help inform the design of progress models in future GPGPU programming frameworks, our work is a compelling example of how programming language semantics and practical formal methods tools (CADP and Alloy) can be brought to bear effectively on an important emerging domain.

\paragraph{Paper structure and auxiliary material}
We start by presenting an illustrative example that is used throughout the paper, and by providing necessary background on GPU programming (\autoref{sec:background}). We then devote a section to each of the contributions described above (\autoref{sec:programming-model}--\autoref{sec:testing}).

Our work focuses principally on cross-vendor inter-workgroup progress; however, we describe preliminary results on testing intra-subgroup behaviors and some specialized progress models provided by the vertically-integrated CUDA framework (\autoref{sec:additional-results}). We conclude with a discussion of related work (\autoref{sec:related}) and conclusions (\autoref{sec:conclusion}).
All software and data for this paper is archived and available, including: the Alloy and CADP models; the progress litmus test suites (in pseudo code and in executable formats for C++, CUDA, Metal, and Vulkan); and the data obtained from running these tests on our various GPU platforms~\cite{artifact}.

\section{Running Example and Background}
\label{sec:background}

\subsection{GPGPU programming frameworks}

There are now many GPGPU programming frameworks, e.g.\ CUDA, OpenCL, Vulkan, and Metal.
We largely use the Vulkan terminology because Vulkan is a well-supported \emph{cross-vendor} framework, which is one of our main focus areas. We note the equivalent CUDA terminology, as CUDA parlance is common in academic literature, and for certain cases where the Vulkan term is cumbersome we use the CUDA term throughout the paper.

The GPGPU concurrency model is hierarchical. A GPGPU program is called a \emph{compute shader} in Vulkan, and a \textit{kernel} in CUDA; we use \emph{kernel} throughout the paper.
A kernel is programmed in a single program multiple data (SPMD) manner:
a single \emph{entry point} function is executed by many \emph{compute invocations} (Vulkan terminology), which we shall refer to more conveniently as \emph{threads} (CUDA terminology).
The threads that execute a kernel are partitioned into \emph{workgroups} of equal, programmer-specified sizes (called \emph{thread blocks} in CUDA). Threads in the same workgroup are generally mapped to the same hardware compute unit (e.g.\ to a single \emph{streaming multiprocessor} on an Nvidia GPU) and thus, can communicate efficiently using fast memory shared across the workgroup.
Threads in the same workgroup are further partitioned into \emph{subgroups} (known as \emph{warps} in CUDA). Threads in the same subgroup are often mapped to the same vector processing unit in which case they can communicate and synchronize extremely efficiently. Different architectures provide different subgroup sizes depending on their architecture.
For example, AMD GPUs feature a subgroup size of 64 (or 32 on newer architectures)~\cite{RDNAWhitepaper}, while Intel GPUs can have subgroup sizes of 8, 16 or 32~\cite[ch. 5]{OneAPI}. Each thread can query ids for their global id (unique per thread), workgroup id, and subgroup id. Threads can additionally query the total number of threads, and the size of workgroups and subgroups.

A well-written GPGPU program will prioritize local interactions (e.g.\ at the subgroup level) over global interactions (e.g.\ across workgroups). To support this, different mechanisms are provided for synchronization between threads at different levels of the execution hierarchy. For example, there are two distinct execution barriers provided as primitives: one is limited to synchronizing threads in the same subgroup; another extends the width of the barrier across the entire workgroup.
It is thus reasonable that threads interacting at different levels of the hierarchy might be subject to different
progress models. The main concern of this work is to detail progress models for interactions of threads across workgroups.

\subsection{Fairness Properties and Semi-fair Progress Models} \label{sec:background-fairness}

Program correctness can be described in terms of \textit{safety properties}, which state that nothing `bad' will happen (commonly expressed as assertions in programs), and \textit{liveness properties}, which state that something `good' will eventually happen.  In concurrent programs, liveness properties are often

\begin{wrapfigure}{r}{6.5cm}
    \centering
    \subfigure[C-style pseudo code for an exchange-lock progress litmus test. \texttt{Exch} atomically stores the argument and returns the value that was at \texttt{m}. Initially \texttt{*m == 0}. \label{fig:mutex-pseudocode}]{
\scriptsize
\begin{tabular}{p{0.45\linewidth} p{0.45\linewidth}}
\toprule
Thread 0 & Thread 1 \\
\midrule
\lstinline|while(Exch(m,1)==1);| &     
\lstinline|while(Exch(m,1)==1);| \\
\lstinline|Store(m,0);| &
\lstinline|Store(m,0);| \\ 
\bottomrule \\[2mm]
\end{tabular}
}
\vspace{.5cm}
\subfigure[State space LTS containing all possible execution paths. \edit{In states where the mutex is acquired, a gray square indicates which thread is currently holding the mutex.} Cycles on states B and E may or may not pose a risk of non-termination, depending on the progress model.\label{fig:mutex-lts}]{
\includegraphics[width=\linewidth]{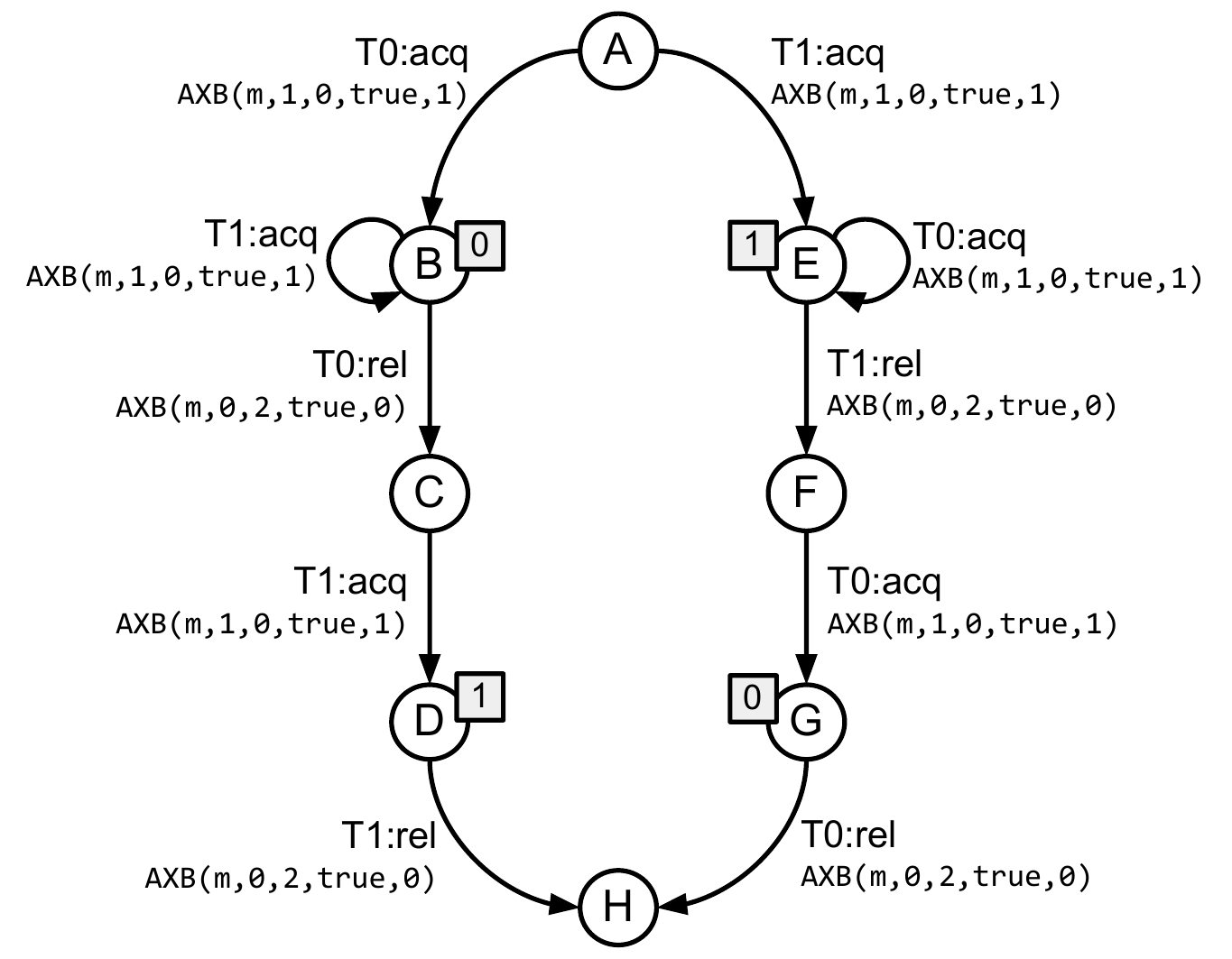}
}
\subfigure[Our DSL with its AXB instruction.\label{fig:mutex-dsl}]{

\footnotesize
\begin{tabular}{p{\linewidth}}
\\[-8mm]
\textit{// Mutex value is at global memory location m,}\\
\textit{// and it is initially set to 0. Program location 2}\\
\textit{// implicitly corresponds to the end of the program}\\
\texttt{Thread 0: [}\\
\quad \texttt{0: AXB(m, 1, 0, true, 1)} \textit{// acquire}\\
\quad \texttt{1: AXB(m, 0, 2, true, 0)} \textit{// release}\\
\texttt{]}\\
\texttt{Thread 1: [}\\
\quad \texttt{0: AXB(m, 1, 0, true, 1)} \textit{// acquire}\\
\quad \texttt{1: AXB(m, 0, 2, true, 0)} \textit{// release}\\
\texttt{]}\\
\end{tabular}
}
 \caption{A mutex progress test, where \texttt{m} is a pointer to an integer mutex, 0 represents unlocked and 1 represents locked. Two threads T0 and T1 contend for acquiring \texttt{m} and then immediately release it.}
    \label{fig:mutex}

    \end{wrapfigure}%
dependent on an assumption that threads are \emph{fairly} executed:
if thread $X$ waits for another thread $Y$ and if $Y$ is \emph{not} fairly scheduled, then $X$ might wait indefinitely, leading a \emph{starvation cycle}, which can cause the program to fail to terminate.

    Throughout the execution of a kernel, threads are said to be \textit{enabled} or \textit{disabled}. In classic concurrency theory, threads can be disabled at various points, e.g.\ while waiting at a mutex. Because GPUs do not have primitive mutexes, the notion of enabled and disabled GPU threads is simple: threads are enabled at the start of kernel execution, and become disabled when they finish kernel execution. GPU frameworks often provide an inter-workgroup barrier primitive. Because our work is largely constrained to inter-workgroup behaviors, we do not consider these instructions (nor their effect on whether a thread is enabled or not) in this work, although we believe this is an interesting avenue for future work.

We say that a scheduler is \emph{unfair} if it nondeterministically chooses \emph{any} enabled thread for execution, regardless of whether other threads are enabled for execution.
Despite a given thread always being ready to make progress, this scheduler can continuously select other threads for execution, leading to the thread being starved.
Classically, there are two main notions of fair schedulers~\cite{DBLP:conf/popl/GabbayPSS80}. 
A scheduler is \textit{weakly fair} if whenever a given thread is continuously enabled, the thread will eventually be scheduled.
A scheduler is \emph{strongly fair} if whenever a given thread becomes enabled infinitely often (without necessarily being continuously enabled), the thread will eventually be scheduled. We refer to these notions of fairness as \emph{(forward) progress models}.

To illustrate the difference between strong and weak fairness, consider a program in which two threads take turns executing, but each execution step by one thread blocks the other thread (this is illustrated in some formulations of the classic ``dining philosophers'' problem). This type of cycle is called a \textit{livelock}. If there exists some interleaving of transitions that would free the threads from their livelock, then strong fairness would ensure that the interleaving would eventually happen; weak fairness provides no such guarantees. Any program that is guaranteed to terminate under weak fairness is also guaranteed to terminate under strong fairness, but not the other way around.

We now show an example to illustrate how to reason about termination for a program under a particular progress model.
The pseudo-code of 
\autoref{fig:mutex-pseudocode} represents a program where two threads compete to acquire and then release a shared mutex. The mutex, represented by a memory location \texttt{m}, is \emph{acquired} by a thread when the thread atomically exchanges the current value of \texttt{m} with 1 and finds that the old value of the mutex was 0. It is \emph{released} by a thread by storing 0 to \texttt{m}.

The state-space of all possible execution paths for the program is shown in \autoref{fig:mutex-lts} in the form of a labelled transition system (LTS). Each transition is labeled by the thread that takes a step, and with \emph{acq} or \emph{rel} depending on whether the step involves acquiring or releasing the mutex.
We discuss the code of \autoref{fig:mutex-dsl}, and the \texttt{AXB} instructions that label the edges of the LTS when we introduce our simple programming language in \autoref{sec:programming-model}.

The LTS contains cycles that indicate potential starvation cycles (i.e.\ non-termination), depending on progress guarantees.
For instance, consider the execution path $A\rightarrow{}B\rightarrow{}B\rightarrow{}B$ \ldots. This is a potential starvation cycle that can occur if Thread 0 (T0) acquires the mutex (action $A\rightarrow{}B$) and then T0 is starved by T1 (the $B\rightarrow{}B$ action repeated indefinitely).

Under a (weakly or strongly) fair scheduler there is no risk of starvation in the mutex example since we can rely on the scheduler to eventually execute T0. This would allow T0 to release the mutex (action $B\rightarrow{}C$) and thus unblock T1, allowing the program to terminate in state $H$.
In contrast, an unfair scheduler offers no such progress guarantees, so that
one thread can acquire the mutex, then the spinning thread can execute indefinitely.

\begin{table}[t]
    \footnotesize
    \centering
        \caption{The semi-fair progress models we consider in this paper.}

    \begin{tabularx}{\linewidth}{L{5cm} L{8.1cm}}
    \toprule
    \textbf{Name} & \textbf{Threads for which fair execution is guaranteed} \\
    \midrule

    OBE \textit{(Occupancy-Bound Execution)} & Threads that have made at least one execution step \\

    HSA \textit{(Heterogeneous System Architecture)} & The thread with the lowest id that has not yet terminated  \\

    LOBE  \textit{(Linear OBE)} \hspace*{4cm} & Threads that have made at least one step, and any thread with a lower id than a thread that has made at least one step\\

    \bottomrule
    \end{tabularx}
    \label{tab:progress-models}
\end{table}

Because cross-vendor GPU programming frameworks do not mention progress guarantees, the safest option is to assume an unfair progress model. This is sufficient for many common GPU use-cases, such as matrix-multiplication, as these applications are embarrassingly parallel and require only infrequent bulk synchronization operations. However, as the diversity of GPU applications increases, so does the need for finer-grained synchronization.

Recent work~\cite{Sorensen-Evrard-Donaldson-18} used linear-time temporal logic to define \textit{semi-fair} progress models that provide more fairness than the unfair progress model, but less fairness than the weak or strong fairness models. These models provide progress on a per-thread basis throughout the execution and are summarized in  \autoref{tab:progress-models}. For each of these progress models, there is a strong and weak variant. In the weak variant, starvation-cycle non-termination behaviors are prohibited. In the strong variant, livelock non-termination behaviors are additionally prohibited. We now provide a brief overview of the models.

Heterogeneous System Architecture (HSA) is a heterogeneous programming framework, with targets that include GPUs~\cite{HSAprogramming}. While cross-vendor support for this framework is limited, the HSA specification provides a clear and intuitive progress model, which states: out of the currently executing threads, the thread with the lowest id is guaranteed eventual execution. Thus, while we do not target the HSA framework directly in our experimental campaign, we consider its progress model an interesting candidate to formalize and test. On the other hand, the \emph{occupancy-bound execution} (OBE) progress model was never officially supported, but proposed in academic works~\cite{GPUBarrierOOPSLA2016, owens-persistent}: it states that a thread that has started executing will continue to be fairly executed. There are many instances of kernels that rely on this progress model~\cite{TPO10,ct08}.

Interestingly, the OBE and HSA progress models are incomparable (as we will show in \autoref{sec:progresslitmustestexamples}). To provide a natural unification of the HSA and OBE models, the \emph{Linear Occupancy-Bound Execution} (LOBE) progress model was proposed~\cite{Sorensen-Evrard-Donaldson-18}. This model guarantees eventual execution to any thread that has taken a step, along with any thread with an id lower than a thread that has taken a step.
Additionally,~\cite{Sorensen-Evrard-Donaldson-18} proposes a unified progress model, named HSA+OBE, which is a direct combination of the HSA and OBE models that has no intuitive mapping to a real design; thus, while we are able to provide executable semantics for this model, we do not consider it in our empirical evaluation.

\subsection{Progress Litmus Test Examples}\label{sec:progresslitmustestexamples}

Here we show two examples of \textit{progress litmus tests}, simple unit tests that characterize various progress properties. These tests have two outcomes: pass (guaranteed termination) and fail (potential non-termination). Tests may fail for two reasons: starvation, in which the scheduler does not give a thread the chance to execute, or livelock, in which a set of threads execute indefinitely, but their execution steps continually block each other. 

\begin{wrapfigure}{T}{5.6cm}
    \centering
  \footnotesize
\begin{tabular}{p{2cm} | p{2.8cm}}
\multicolumn{2}{c}{\textbf{one-way ProdCons (increasing id)}} \\
\toprule

Thread 0 & Thread 1 \\
\midrule
\lstinline|Store(flag,1);|  &     
\lstinline|while(Load(flag==0));|
\\ 
\bottomrule
\end{tabular}
 \caption{A one-way (only 1 thread spins) producer-consumer (ProdCons) progress litmus test. 
 \edit{The test is described as ``increasing id'' because the thread with the higher id waits on the thread with the lower id.}
 }
    \label{fig:producer-consumer-early}

\end{wrapfigure}
The behavior is determined with respect to a given progress model. \edit{\autoref{fig:producer-consumer-early} shows a producer-consumer idiom where T1 waits for T0 to write a flag.} Under an unfair progress model, this test might not terminate, as T1 might execute indefinitely, while T0 starves. In a fair progress model (strong or weak), T0 is guaranteed to eventually execute, and thus T1 will eventually be unblocked, and thus, this test is guaranteed to terminate (i.e.\ it passes) under a fair progress model. Our second example is the mutex progress litmus test shown in \autoref{fig:mutex}, which we analyzed similarly in \autoref{sec:background-fairness}.

While these two tests behave similarly under classic progress models, they behave differently under semi-fair schedulers. One such scheduler, OBE, states that once a thread starts executing, i.e.\ has executed an instruction, it will continue to be fairly executed. The producer-consumer test fails under this model as T0 is not guaranteed to be fairly scheduled until it starts executing, thus T1 might spin indefinitely. However, the mutex test will pass, because any thread that has acquired the mutex will be fairly scheduled, and thus will eventually release it. The two tests have the opposite behaviors under HSA: The producer-consumer test is guaranteed to terminate under HSA as T0 is guaranteed progress, which will unblock T1. \edit{However, the mutex test is not guaranteed to terminate under HSA, as T1 could acquire the mutex, and then starve while T0 spins.} This is because HSA does not guarantee eventual execution to any thread except the thread with the lowest id that has not terminated.

\subsection{A Motivating Applications Use-case}

\edit{In \autoref{sec:intro}, we reference several examples where existing GPU programs use progress model assumptions to achieve high performance that otherwise would not have been possible. We will discuss one such example in more detail here: the GPU graph applications presented in~\cite{Sorensen-Pai-Donaldson-19}.}

\edit{This work discusses an OpenCL backend for IrGL, a DSL for GPU graph processing.
IrGL can be used to express many common graph algorithms, including breadth-first search, single-source shortest path, and PageRank.

The DSL has a \textit{vertex-centric} approach in which all graph vertices are processed in parallel.
Each vertex either pulls in updates from its neighbors, or pushes out updates to its neighbors. A complete pass through all the vertices is known as an \textit{epoch}. The algorithms iterate over epochs until a fixed point is reached, i.e.\ an epoch is computed in which no vertices are updated. Global synchronization is required between each epoch as the vertex computation requires up-to-date values from the previous epoch. This approach to graph computation has been shown to be competitive on both CPUs~\cite{beamer2017gap} and GPUs~\cite{Gunrock}.}

\edit{The OpenCL compiler for IrGL can be configured to make \textit{no} assumptions about the GPU progress model, in which case, the global synchronization between epochs must be implemented via rendezvous with the CPU host, i.e.\ by ending the GPU kernel every epoch, and relaunching the GPU kernel from the CPU, until the fixed point is reached. This rendezvous with the CPU (and as such, through the GPU driver) can be costly. To address this, the compiler allows the user to specify that they wish to make (officially unsupported) assumptions about the progress model. Under this direction, the compiler will emit code that uses a limited global synchronization barrier (as described in~\cite{XF10, GPUBarrierOOPSLA2016}), and thus the synchronization between epochs can occur entirely on the GPU, without cross-device rendezvous.}

\edit{This optimization, enabled by the assumption of an unsupported progress model, was shown to have significant performance improvements. The paper discusses a large evaluation consisting of 6 GPUs (from Nvidia, Intel, AMD, and ARM), 17 graph applications, and 3 idiomatic graph inputs: a total of 306 benchmarks. Across these benchmarks, the global barrier optimization achieved a statistically significant speedup in 57\% of the benchmarks. In 20\% of the cases, the speedup was over $2\times$. At the extreme end, GPUs from AMD and Intel both had benchmarks for which the optimization provided over a $10\times$ speedup. In these cases, the graphs were high diameter (requiring many epochs) with very small amounts of computation per epoch.}

\edit{While we do not extend our reasoning to full application use-cases as described here, this section shows that progress model assumptions (and hopefully someday official support) can enable significant performance in fundamental applications.}

\section{Programming Language \label{sec:programming-model}}

Our eventual goals in this work are two-fold: using model checking as a termination oracle for progress litmus tests under various progress models, and for synthesis of progress litmus test cases. In both settings, scalability is a known issue. Thus, we strive to provide the simplest possible programming language that allows us to capture interesting progress litmus tests. To do this, we define a simple \texttt{goto} language with a single atomic instruction. Later, in \autoref{sec:methodology} we will show how to compile this language into executable programs in various GPGPU frameworks.

\paragraph{Global Memory}
Our model of memory is a shared monolithic mapping from locations (of type \textbf{\texttt{loc}}) to values (of type \textbf{\texttt{val}}). Memory locations can be dereferenced, similar to C-style pointers. Although GPUs have hierarchical memory regions, our goal is to model inter-workgroup interactions. Thus, this memory region will correspond to the global memory region on GPUs, shared across all threads.

\paragraph{Threads} A thread is described by a sequence of instructions. Thread-local storage consists only of a program counter (\texttt{pc}) that determines the next instruction to execute. We opt not to include any thread-local registers (or variables) to aid in the scalability of our test-case synthesis (\autoref{sec:alloy}).

\paragraph{Instructions}

There is a single kind of instruction called \atomicxb{}, or \axb for short.
An \axb instruction consists of:
a location \texttt{checkLoc} (of type \textbf{\texttt{loc}}) to be checked,
a value \texttt{checkVal} (of type \textbf{\texttt{val}}) against which the contents of \texttt{checkLoc} should be compared, and an integer program location \texttt{jumpInst} to which control should jump if the compared values are equal. An instruction also includes a boolean, \texttt{doExch}. If the value of \texttt{doExch} is \emph{true} then the instruction causes the contents of \texttt{checkLoc} to be replaced with a new value, \texttt{exchVal} (of type \textbf{\texttt{val}}), which also forms part of the instruction. If \texttt{doExch} is \emph{false} then the \texttt{exchVal} component is unused.

The semantics of an \axb instruction is given by the following C-like pseudocode, executed in an atomic fashion, where \texttt{thread} refers to the state of the calling thread:

\noindent
\begin{minipage}{\textwidth}
\begin{lstlisting}[basicstyle=\scriptsize\ttfamily, morekeywords={loc, val, bool, inst}]
 void ATOMIC_EXCH_BRANCH(loc checkLoc, val checkVal, int jumpInst, bool doExch, val exchVal) {
   thread.pc = (*checkLoc == checkVal) ? jumpInst : thread.pc + 1;
   if (doExch) *checkLoc = exchVal; 
 }
\end{lstlisting}
\end{minipage}

Intuitively, the \axb instruction combines the functionality of an atomic exchange and a conditional branch. There is no reason to return any value as our language does not provide any thread-local storage to manage a return value. 

While this language may seem simple, it is expressive enough to capture many interesting progress litmus tests. It can mimic a plain memory store (if \texttt{jumpInst} is set to refer to the next instruction in program order). It can perform an atomic exchange, necessary for a spin-lock.
For example, \autoref{fig:mutex-dsl} illustrates our mutex example using \axb. Our language can perform a spin-loop on a plain load, e.g.\ as seen in the producer consumer test of \autoref{fig:producer-consumer-early}, when \verb|doExch| is set to false.

For ease of exposition, for the remainder of the paper, we provide our progress litmus tests using C-style syntax, including: \texttt{if} statements, \texttt{goto} statements, atomic exchange instructions (\texttt{Exch}), and memory accesses (using ``*'' to dereference the memory location). We use this presentation with the understanding that \axb is expressive enough to capture these behaviors, and that the model checking and programming synthesis is built on this single, yet expressive, instruction. 

\section{Executable Semantics of Progress Models}
\label{sec:cadp}

The programming language of \autoref{sec:programming-model} lets us write progress litmus tests, i.e. small concurrent programs that are guaranteed to terminate under strong fairness, but might not terminate in the absence of progress guarantees.
Now we are interested in checking whether these tests pass---are guaranteed to terminate---\emph{with respect to a given progress model.}
We achieve this by expressing progress models through executable semantics and using model checking to look for potential non-termination.
These executable semantics are written in the LNT formal language~\cite{Champelovier-Clerc-Garavel-et-al-10-v6.7},
which has associated labeled transition system (LTS) semantics---for a formal treatment of LTSs see~\cite{BaierKatoen08}.
Verification is done with the CADP tool suite,
which can check properties of the LTS expanded from an LNT specification,
where properties are expressed using \textit{Model Checking Language} MCL~\cite{MCL-Mateescu-Thivolle-08}.
The LNT and MCL code excerpts presented here are slightly edited for ease of exposition.

\subsection{Progress Model Specification}
Our LNT specification is made up of two parts: (1)~an interpreter for our programming language, which takes as input a list of \axb instructions for each thread; and (2)~a progress model.
The progress model monitors each thread's execution steps and whether the thread has terminated or not (i.e.\ the progress model is signaled when a thread has completed execution of its own program), and at each step it must report the current set $F$ of threads for which fair execution, i.e.\ eventual execution, is guaranteed.
So in our framework, a single execution step contains the following information: (a) the id of the thread taking this step, (b) the \axb instruction that it executes, and (c) the set $F$ of thread ids that are guaranteed fair execution at the point before this step is executed.
At each step, a progress model monitors (a) and (b), and is responsible for defining (c).

\begin{figure}[t]
\begin{lstlisting}[language=LNT,xleftmargin=.25in,basicstyle=\scriptsize\ttfamily]
process OBE [Step: ExecutionStep, Terminate: Natural] is
  var
    tid: Nat,     -- thread ID
    axb: AxbInst, -- AXB instruction
    F:   NatSet   -- set of threads guaranteed fair execution
  in
    F := {}; -- At the beginning, no thread is guaranteed fair execution
    loop
      select -- non-deterministic choice operator
        Step(?tid, ?axb, F); -- thread tid executes an instruction
        F := insert(tid, F)  -- thread tid is now guaranteed fair execution
      []
        Terminate(?tid);     -- thread tid has terminated its own program
        F := remove(tid, F)  -- thread tid will not be executed anymore
      end select
    end loop
  end var
end process
\end{lstlisting}
\caption{Specification of the OBE progress model in the LNT formal language.}
\label{fig:lnt-obe}
\end{figure}

\begin{figure}[t]
\includegraphics[width=0.7\linewidth]{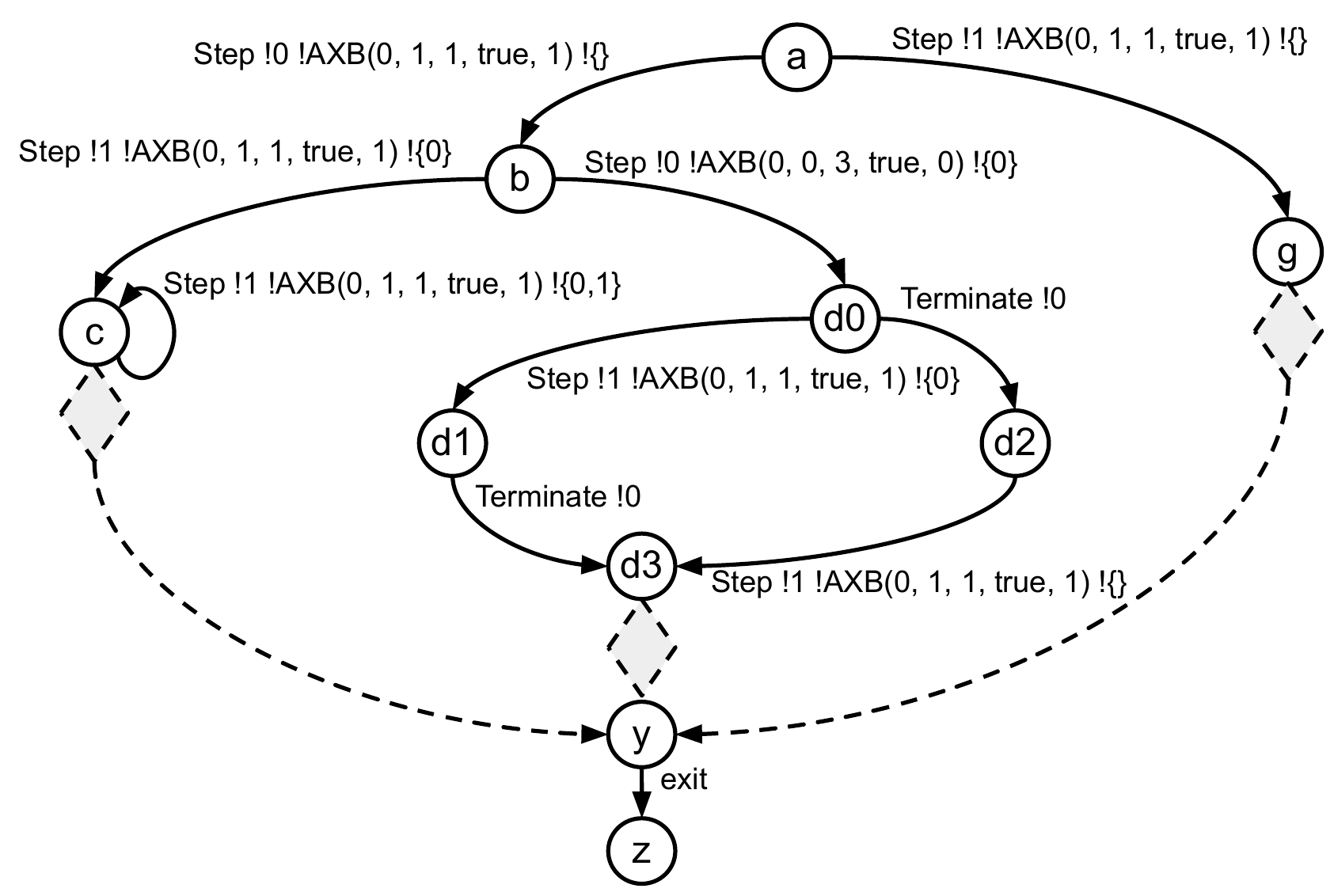}
\caption{\edit{A portion of the full LTS generated by CADP from the LNT model of our mutex example (\autoref{fig:mutex}) with the OBE progress model. Transitions can either be a \texttt{Step}, where a thread executes an \texttt{AXB} instruction, or a \texttt{Terminate}, where a thread ends execution. The step label syntax is defined in \autoref{sec:cadp-checking}. Because the complete LTS is too big to be presented here, some parts are abstracted with gray diamonds and dotted arrows. We highlight the path where thread 0 obtains the mutex ($a\xrightarrow{}b$). The actions between states $d0$, $d1$, $d2$ and $d3$ represent the concurrent termination of T0 with the mutex acquisition by T1. The LTS ends with a special \texttt{exit} action generated by CADP.}}
\label{fig:lnt2lts}
\end{figure}

Figure~\ref{fig:lnt-obe} shows the specification of the OBE semi-fair progress model that we introduced in \autoref{sec:background-fairness}
and \autoref{tab:progress-models}, which guarantees fair execution to threads that have made at least one execution step.
\texttt{Step} and \texttt{Terminate} are synchronization primitives between the progress model and the language interpreter (specified by another LNT process and not presented here).
On \texttt{Step} (line~10), the progress model receives (operator \lstinline[language=LNT]|?|) the thread id and the \axb instruction (the specific \axb instruction being executed is not relevant for the OBE progress model, so we don't make use of the value received in the \lstinline[language=LNT]|axb| local variable).

Moreover, the OBE progress model provides $F$, the set of threads that are guaranteed fair execution before this step.

After the execution step (line~11), the semantics of OBE are implemented by adding the thread id to $F$.
On \texttt{Terminate} (line~13), the progress model receives the terminating thread's id, and then (line~14) updates $F$ accordingly.

Using a similar approach, we wrote progress model specifications for the 6 progress models

we found in prior work, namely: \textit{unfair} (no thread is ever guaranteed fair execution), \textit{fair} (all threads are always guaranteed fair execution until they terminate), HSA, OBE and LOBE (see Table~\ref{tab:progress-models}). We also wrote a progress model for the HSA+OBE model of prior work. This model states that a thread is guaranteed fair execution if it has the lowest id of non-terminated threads, or if it has made at least one execution step. However, its original presentation admits that this model exists as a logical exercise and does not have an intuitive implementation. Thus we do not consider it further in this work, i.e.\ to characterize our progress litmus tests or empirical results.

\subsection{Checking Whether a Test Passes or Fails Under a Progress Model \label{sec:cadp-checking}}

\begin{figure}[t]
\centering
\subfigure[Mutex under HSA.\label{fig:lts-hsa}]{
\includegraphics[width=0.33\linewidth]{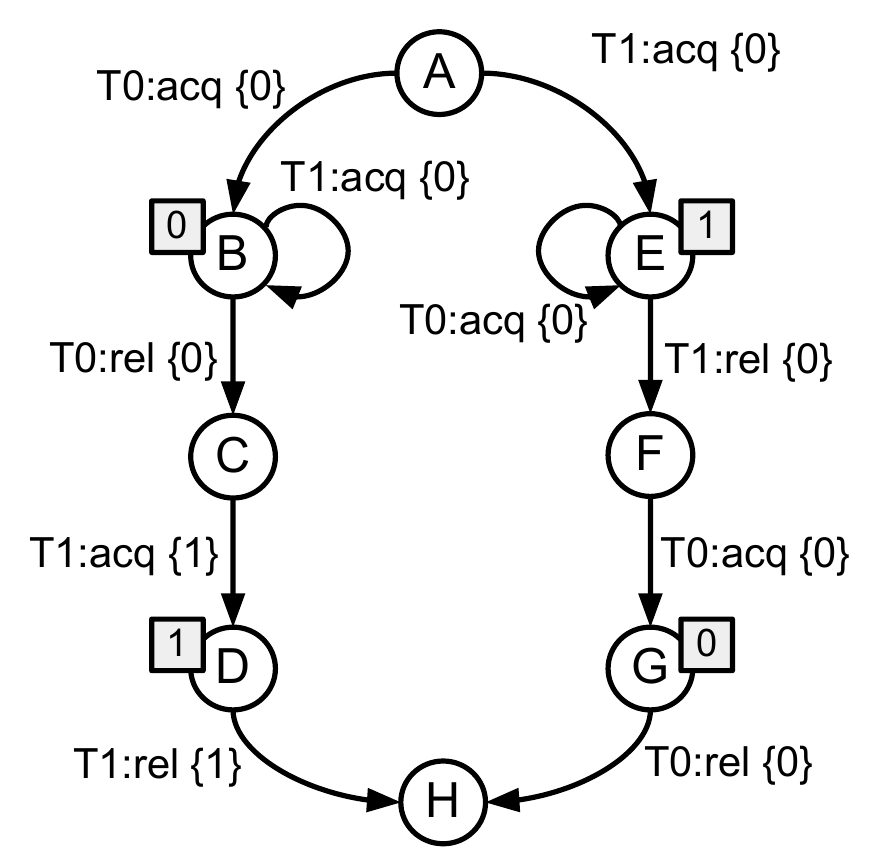}
}
\subfigure[Mutex under OBE.\label{fig:lts-obe}]{
\includegraphics[width=0.42\linewidth]{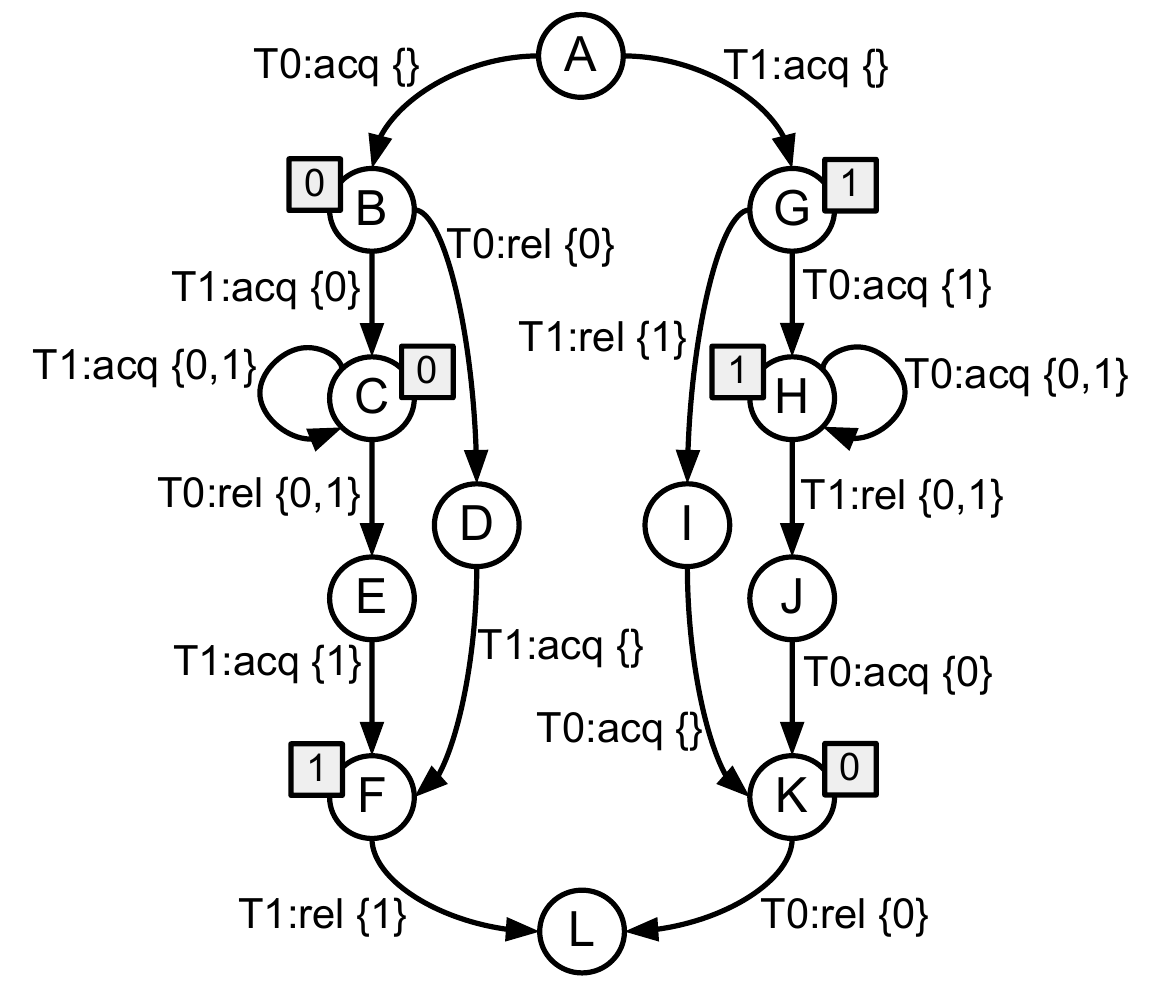}
}
\subfigure[Livelock sensitive to weak/strong fairness.\label{fig:lts-weakstrong}]{
\includegraphics[width=0.18\linewidth]{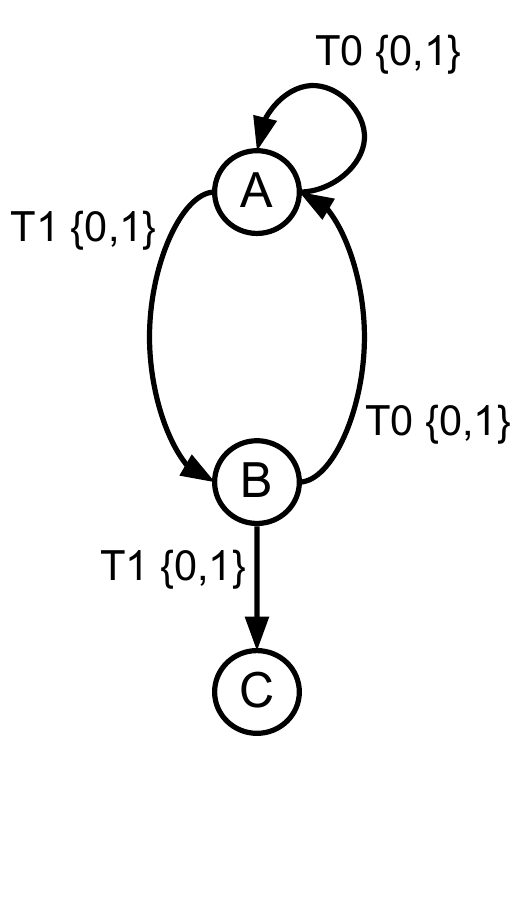}
}
    \caption{Abstracted LTSs from our executable semantics for our mutex example, where only \texttt{Step} transitions are showed and abbreviated by its thread id, instruction shorthand (\texttt{acq/rel}) and \edit{its set $F$ of fairly executed threads (noted with thread ids inside braces, with ``$\{\}$'' denoting the empty set)}. \ref{fig:lts-weakstrong} illustrates a livelock cycle that is sensitive to weak and strong fairness: the T1 transition that escapes the cycle is guaranteed to be eventually taken under strong fairness, but not under weak fairness.}
    \label{fig:lts-hsa-obe-weakstrong}
\end{figure}

To determine whether a test passes (i.e.\ is guaranteed to terminate) we check for the absence of non-termination cycles in the LTS.
Depending on the progress model, not all cycles correspond to potential non-termination.
Intuitively, a cycle that may lead to non-termination is one that \emph{fairly executed threads are not guaranteed to escape.}

CADP can explore the LTS it generates from our LNT specification of a litmus test under a given progress model.
\edit{For instance, \autoref{fig:lnt2lts} illustrates parts of the LTS obtained for our mutex example with the OBE progress model.}
These LTSs have three kinds of transitions: \texttt{Step} and \texttt{Terminate} transitions, which represents a thread executing an instruction and terminating, respectively,
and a special \texttt{exit} transition that CADP generates to indicate that all threads have terminated.

\edit{In order to clarify the presentation of our verification approach, \autoref{fig:lts-hsa-obe-weakstrong} presents abstracted LTSs for our mutex example under the HSA and OBE progress models, where only \texttt{Step} transitions are kept and renamed.}
\texttt{Terminate} and \texttt{exit} transitions are omitted \edit{(to be precise, \texttt{Terminate} actions are coalesced with the previous \texttt{Step} of the terminating thread)} since they are never present inside a cycle, as the termination of a thread or of all threads is not something that can be repeated.
Instead, we focus on the \texttt{Step} transitions, which in the figure are abbreviated by their thread id, instruction shorthand and $F$ components, e.g. ``\lstinline[language=LNT]|Step !0 !AXB(m, 1, 0, true, 1) !{1})|'' \edit{(which reads as: ``a step by thread 0, executing the \texttt{AXB(m, 1, 0, true, 1)} instruction, while $F$ contains thread 1'')} is abbreviated to ``\lstinline[language=LNT]|T0:acq {1}|.''
We can first observe that different progress models can lead to different shapes of LTSs, due to the fact that they set $F$ in different ways depending on which steps have been executed so far.

Let's focus on the cycle that happens when T0 loops on trying to acquire the mutex when it is held by T1.
Only a step from T1 can release the mutex and thus escape the cycle.
Now, whether T1 is guaranteed fair execution or not depends on the progress model.
Under HSA (Figure~\ref{fig:lts-hsa}) this cycle is on state E.
In both outgoing transitions from E, the set $F$ of fairly-scheduled threads is $\{0\}$, i.e.\ T1 \emph{is not} guaranteed fair execution.
Under OBE (Figure~\ref{fig:lts-obe}) this cycle is on state H, and we have $F=\{0, 1\}$ for both outgoing transitions so that T1 \emph{is} guaranteed fair execution.
Accordingly, this cycle represents non-termination (by starvation of T1) under HSA, but not under OBE.

In general, whether a cycle indicates potential non-termination is also influenced by weak vs.\ strong fairness (see \autoref{sec:background-fairness}).
Consider the LTS in Figure~\ref{fig:lts-weakstrong} under a progress model that guarantees fair execution of both T0 and T1.
Here we focus on the livelock cycle $A\rightarrow{}B\rightarrow{}A\rightarrow{}B\rightarrow{}...$ where both T0 and T1 make progress.
To break this cycle, we need more than eventual execution of T0 and T1: we need T1 to take two steps in a row; this breaks the livelock cycle and allows the test to terminate.
While threads loop in this cycle, the T1 transition that escapes the cycle can \emph{infinitely often} be taken $B$, so under strong fairness we have the guarantee that it will eventually be taken, and thus the cycle cannot cause a non-terminating livelock.
However, the escaping transition cannot be taken from $A$, so while threads loop the cycle it is not \emph{always} possible to escape, and therefore the escaping transition is not guaranteed to be eventually taken under weak fairness.
This illustrates how, for the same progress model, a livelock cycle can represent non-termination under weak fairness but not under strong fairness.

\noindent\begin{minipage}{.37\textwidth}
\begin{lstlisting}[caption={MCL formula to match non-termination cycles under weak fairness.},label=lst:weak,frame=tlrb,language=MCL,basicstyle=\ttfamily\scriptsize]
< true* . {Step ... ?F:NatSet
            where is_in(tid,F)} >
 < for tid:Nat from 0 to MAX_ID do
     if is_in(tid, F) then
       true* . {Step !tid ...}
     end if
   end for
 > @
\end{lstlisting}
\end{minipage}\hfill
\begin{minipage}{.59\textwidth}
\begin{lstlisting}[caption={MCL formula to check for absence of non-termination cycles under strong fairness.},label=lst:strong,frame=tlrb,language=MCL,basicstyle=\ttfamily\scriptsize]
[ (not "exit")* ]
 < ( {Step ?tid:Nat ... ?F:NatSet where is_in(tid,F)}
     or {Terminate ...}
   )* . (
     "exit"
     or {Step ... ?F:NatSet where is_empty(F)}
   )
 > true
\end{lstlisting}
\end{minipage}

%%% MCL primer
We check the presence or absence of non-termination cycles using MCL formulas.
MCL is an expressive value-passing modal $\mu$-calculus~\cite{Kozen83} language, and we only describe the subset we use (see~\cite{mcl4ManPage} for more details).
To match paths in the LTS, MCL offers operators to match a transition's label, and on top of that operators to match paths made of such transitions.

At the label level: \lstinline[language=MCL]|true| matches any label, \lstinline[language=MCL]|"foobar"| matches exactly ``\textit{foobar}'', and
\lstinline[language=MCL]|"foo" or "bar"| matches either ``\textit{foo}'' or ``\textit{bar}''.
Curly brackets do regular-expression matching on a label, with the possibility to skip over some parts of the label and to capture data value, e.g. \lstinline[language=MCL]|{Foo ... ?n:Nat}| would match ``\textit{Foo !bar !baz !42}'', skipping ``\textit{!bar !baz}'' and capturing the natural value 42 in variable \lstinline[language=MCL]|n|.
Moreover, such matching can be conditioned with a \lstinline[language=MCL]|where| clause, e.g.\ \lstinline[language=MCL]|{?n:Nat where n < 10}| would not match ``\textit{42}''.

At the path level: transitions can be concatenated with ``\lstinline[language=MCL]|.|'', grouped with parentheses, and repeated with ``\lstinline[language=MCL]|*|'', such that e.g.\ \lstinline[language=MCL]|"a".("b"."c")*| matches a path made of transitions labelled ``\textit{a}, \textit{b}, \textit{c}, \textit{b}, \textit{c}, \dots''.
At the top level, we can check ``\textit{for all paths}'' with \lstinline[language=MCL]|[ ]|, ``\textit{there exists a path}'' with \lstinline[language=MCL]|< >|, and ``\textit{the path repeats infinitely in a cycle}'' with \lstinline[language=MCL]|< > @|.
The paths inside these operator can be constructed with \lstinline[language=MCL]|for| loops and \lstinline[language=MCL]|if| conditionals, e.g.:

\smallskip
\lstinline[language=MCL]|< for n:Nat from 0 to 4 do if is_odd(n) then {Foo !n} end if end for >|

\smallskip
\noindent is equivalent to: \lstinline[language=MCL]|< {Foo !1} . {Foo !3} >|.

\paragraph{Weak fairness:} we detect non-termination by looking for livelock cycles where all fairly executed threads make progress.
The progress model guarantees that all these threads should keep on being scheduled, and that is the case in such cycles.
Even if there exist transitions of fairly executed threads that escape the livelock cycle, weak fairness does not guarantee that they will eventually be taken.
To check for these livelock cycles, we use the formula of \autoref{lst:weak} that can be read as:
there exists a path (\lstinline[language=MCL]{<  >}) that starts with zero or more transitions of any kind (\lstinline[language=MCL]{true*}) followed by a \texttt{Step} transition where we capture a non-empty set $F$ of threads which are currently guaranteed fair execution (\lstinline[language=MCL]|{Step ... ?F:NatSet where not(is_empty(F))}|).
This path is followed by a cycle (\lstinline[language=MCL]{<  > @}) where each thread that is in $F$ takes at least one step, with zero or more transitions of any kind between these steps (the \lstinline[language=MCL]|true* . {Step !tid ...}| inside the \lstinline[language=MCL]{for} loop).

We note that this formula assumes that $F$ stays the same during the cycle.
To ensure this, we request progress models to not remove threads from $F$ until they terminate, which is the case for all progress models we consider.
With this hypothesis, in a cycle, no thread can be removed then added back in $F$, and therefore $F$ is stable over a cycle.
Also, we notice that the \lstinline[language=MCL]{for} loop imposes an order on thread ids in the livelock cycle transitions, yet this is not an issue here since the \lstinline[language=MCL]{< > @} operator matches over the ``unrolled path'' of a cycle.
For instance, consider a cycle made of transitions labelled by thread ids 1, 3, 2, in this order.
If all three threads are in $F$, then \lstinline[language=MCL]|< true*.1.true*.2.true*.3 >@| matches the cycle's unrolled path: \textbf{1}, 3, \textbf{2}, 1, \textbf{3}, 2, 1, \ldots

\paragraph{Strong fairness:} we detect the absence of non-termination cycles by checking that from any state, there exists a path made of steps by fairly executed threads that reaches either termination of the whole program, or a state where no thread is guaranteed fair execution.
In particular, from states visited infinitely often by a cycle, if there is a step by a fairly executed thread that can escape the cycle, then this cycle does not represent non-termination since strong fairness guarantees that this escaping step will eventually be executed.
Note that paths of steps by fairly executed threads may reach states where no thread is guaranteed fair execution, but not all threads are terminated: this is for instance the case of states A, D and I in \autoref{fig:lts-obe}.
In such states, given that there is still work to do, at least one thread is eventually scheduled to take an execution step.

To check for such paths, we use the formula of \autoref{lst:strong} that can be read as:
for all paths (\lstinline[language=MCL]|[ ]|)
made of zero or more transitions that are not the final transition (\lstinline[language=MCL]|(not "exit")*|)---i.e.\ from all non-final states---there exist a path (\lstinline[language=MCL]|< >|) made of:
zero or more (\lstinline[language=MCL]|( )*|) transitions which are either:
a step by a fairly executed thread (\lstinline[language=MCL]|{Step ?tid:Nat ... ?F:NatSet where is_in(tid, F)}|),
or the termination of a thread (\lstinline[language=MCL]|{Terminate ...}|);
followed by a single transition which is either:
the termination of all threads (\lstinline[language=MCL]|"exit"|),
or a step where no thread is guaranteed fair execution (\lstinline[language=MCL]|{Step ... ?F:NatSet where is_empty(F)}|).

\section{Automatic Synthesis of Progress Litmus Tests}
\label{sec:alloy}

We now turn our attention to test-case synthesis for progress litmus tests. To do this, we encode the semantics of our programming language in the Alloy modeling framework~\cite{alloy} as LTSs. We then develop constraints that characterise the LTSs that we are interested in: those that are well-formed,
guaranteed to terminate under a fair progress model,
and prone to non-termination under an unfair progress model.
We invoke Alloy in an iterative manner, following \citet{lustig+17} and \citet{chong+18}, with the aim of finding as many LTSs as possible that satisfy those constraints. From the LTS, we can derive a concurrent program, i.e.\ a progress litmus test.

\subsection{LTS Constraints for Synthesis}

As a foundation, we encode the notion of a program to consist of $N$ distinct sub programs, where $N$ is the total number of threads. Each sub program is a sequence of \axb instructions. The global memory state is implemented as a relation from a memory location to a value. Thread local states are implemented as $N$ program counters, each of which point to an instruction in a program. A state contains one global memory and $N$ local states. We implement the semantics of \axb as a state transition relation, which updates the global and local state. 

The following constraints ensure that only well-formed LTSs are synthesized, i.e.\ LTSs that correspond to concurrent programs that are guaranteed to terminate under strong fairness:

\begin{itemize}
    \item \textbf{Start state:} There is a unique `start' state in the LTS, and every state is reachable from it. Memory is initialized to 0 and thread program counters are initialized to the first instruction of their respective program.

    \item \textbf{Always enabled:} At every state, if a thread $t$ has not terminated, i.e.\ the program counter points to a valid \axb instruction $i$, then there exists an action out of that state that corresponds to $t$ executing $i$. Our programming language does not include the ability for threads to be disabled, and thus, our synthesis should not generate tests that require this feature either.
    
    \item \textbf{End state:} There is at least one distinguished `end' state in the LTS. An end state represents the successful termination of the program. No transitions are possible from an end state. 
    
    \item \textbf{Possible termination:} From every non-end state in the LTS, there exists a path to an end state. This ensures that termination \textit{is possible} from any point in the LTSs, but it \textit{may not be guaranteed} depending on the progress model.

\end{itemize} 
While the above constraints describe a valid LTS, we would like to further specialize our synthesis to generate \textit{interesting} tests. For example, the current constraints would synthesize a program consisting of a single thread, executing a single \axb instruction. This is a valid program but it does not yield any insight into progress models. 

To synthesize interesting tests, we encode the following constraint:
\begin{itemize}

    \item \textbf{Non-termination cycle:} There must be a cycle in the LTS corresponding to a possibility of non-termination. Combined with the \textbf{possible termination} constraint above, this generates tests that always have a chance to terminate (i.e.\ there is always a path to the end state), but the guarantee to escape the non-termination cycle will depend on the progress model. 
    
\end{itemize}

\subsection{LTS Minimality Constraints}

The above constraints synthesize \textit{interesting} tests, but the tests are not guaranteed to \textit{non-redundant}. That is, we could imagine synthesizing an interesting test $t$, then synthesizing another test $t'$ which is exactly the same as $t$, except it adds an extra \axb instruction that targets a new memory location.
Test $t'$ cannot characterize any meaningful progress model features additional to those characterized by $t$, and thus $t'$ is redundant.
Constraints that avoid redundant tests such as $t'$ are called \textit{minimality} constraints. 

While prior works have precisely encoded minimality constraints~\cite{lustig+17}, we were unable to derive practical minimality contraints for this domain; any constraints we tried to implement completely stalled the synthesis process. Instead, we rely on three heuristics, that were developed through iterations of manual inspection and abstraction. 

\begin{itemize}
    \item \textbf{Inter-thread influence:} A store from one thread must be loaded by and influence the control flow of another thread. This constraint ensures that tests that differ only by thread-local behavior (and thus do not say anything meaningful about progress models) are not generated. 
    \item \textbf{All branches possible:} There must exist executions through the LTS that take both branches of every conditional. This constraint removes tests that have differ only by control flow targets in dead branches.
    \item \textbf{Meaningful comparisons:} for some \axb instructions, the branch target is the same as the next instruction (in which case the \axb instruction mimics a vanilla store instruction). In such cases, we restrict the comparison to only one value; otherwise redundant tests could be generated that simply change the comparison value, while the control flow (i.e.\ paths through the LTS) remains the same. 
\end{itemize}

\subsection{Synthesis Example}

\edit{To illustrate our synthesis constraints with a concrete example, we now show how the mutex progress test (\autoref{fig:mutex}) satisfies each of the constraints, and thus, could be automatically generated by our approach.}
\begin{itemize}
    \item \edit{\textbf{Start state (\textit{well-formed})}: As shown in the LTS, there is a unique start state, \texttt{A}.}
    
    \item \edit{\textbf{Always enabled (\textit{well-formed})}: It is always possible to execute a thread until it terminates. Notice in the LTS that there are two paths out of states \texttt{A}, \texttt{B}, and \texttt{E}, corresponding to executing thread 0 or 1. All other states represent points in the execution where one of the threads has terminated.}
    
    \item \edit{\textbf{End state (\textit{well-formed})}: In this case, there is a single end state: \texttt{H}; however, it is possible to have more than one. }
    
    \item \edit{\textbf{Possible termination (\textit{well-formed})}: From every state, except for the end state \texttt{H}, there is a path to the end state.}
    
    \item \edit{\textbf{Non-termination cycle (\textit{interesting})}: Examining the LTS, there are two non-termination cycles: \texttt{B}$\xrightarrow{}$\texttt{B} and \texttt{E}$\xrightarrow{}$\texttt{E}. These correspond to the situation where one thread is spinning, waiting for the other to release the mutex.}
    
     \item \edit{\textbf{Inter-thread influence (\textit{minimality})}: We will describe the inter-thread influence (i.e.\ how stores from one thread influence the other thread) by showing how the stores in thread 0 influence thread 1. The reasoning for how thread 1 influences thread 0 is similar. In thread 0, both program instructions perform a store to the mutex: instruction 0 stores a 1 and instruction 1 stores a 0. In thread 1, program instruction 0 performs a branch depending on whether the mutex value is 0 or 1. Thus, both stores of thread 0 can influence the control flow of thread 1.}
     
    \item \edit{\textbf{All branches possible  (\textit{minimality})}: When examining the \texttt{AXB} code, consider the acquire instructions (at program location 0 for both threads). This constraint states that both branches specified in these instructions must be possible in an execution, i.e.\ we do not allow unreachable branches in our tests. From the LTS, we can see that it is possible for both branches to be taken: for example, in the case of thread 1, the branch when the comparison succeeds (i.e.\ the mutex acquire fails), which occurs in the action: \texttt{B}$\xrightarrow{}$\texttt{B}. The other branch occurs when the comparison fails (i.e.\ the mutex acquire succeeds) occurs in the action: \texttt{C}$\xrightarrow{}$\texttt{D}.}
    
    \item \edit{\textbf{Meaningful comparisons  (\textit{minimality})}: When examining the \texttt{AXB} code, notice the release instructions (at program location 1 for both threads) do not have meaningful control flow, i.e.\ the branch location is the same as the next instruction in program order. Thus, the comparison value (the argument at index 1) is constrained to be 0. Otherwise the synthesis would generate redundant tests, changing only this comparison value. }
\end{itemize}

\subsection{Synthesis Results}
We now report on our experience using the Alloy analyzer to synthesize progress litmus tests. We performed 5 different synthesis runs, each configured so that tests would feature a particular number of threads and total number of instructions (across all threads). Each synthesis run was allowed to execute for 7 days on a machine with a 20-core Intel Xeon Gold 6230 CPU (2.10 GHz), with 1 TB of RAM. Our synthesis runs were performed in sequence so that each run could have sole access to the machine's resources.

\begin{wraptable}{R}{6cm}
    \centering
    \scriptsize
    \caption{Overview of the 5 synthesis runs: each row shows the bounds of each run and how many tests were synthesized; they differ only by the number of threads and the total number of instructions (Instrs), distributed across all threads. The Actions and States are upper-bounds, while the threads and instructions are exact constraints.}
    \begin{tabular}{r r r r | r r}
    \toprule
         \textbf{Threads} & \textbf{Instrs} & \textbf{Actions} & \textbf{States} & \textbf{\# Tests}  \\
         \midrule
         2 & 2 & 8 & 8 & 8 \\ 
         2 & 3 & 14 & 12 & 176\\
         2 & 4 & 16 & 24 & 173\\
         3 & 3 & 16 & 24 & 21\\
         3 & 4 & 16 & 24 & 105\\
         \bottomrule
    \end{tabular}
    
    \label{tab:test-generation-results}
\end{wraptable}
\autoref{tab:test-generation-results} shows the test synthesis bounds on each of our 5 runs. Our synthesis constraint requires cycles consisting of at least two threads, and each thread must execute at least one instruction in the cycle, thus the smallest configuration it was possibly to use for synthesis required 2 threads and 2 instructions.
The largest configuration for which synthesis was feasible required 3 threads and 4 instructions; pilot runs with higher bounds did not produce any tests, even though they ran for longer than 24 hours. All synthesis runs are constrained to 2 memory locations and consider only boolean values.

The number of actions and states were selected based on a conservative upper-bound computed as follows: Given each memory location has a $0/1$ value, and the program counter can range from zero to the number of instruction per thread plus one, we multiply the possible values of each to obtain the bound on states (which is tight). Then, for each state we consider that each non-terminated thread contributes one outgoing action, giving us a bound on actions as well. We were able to use this upper-bound for the first three configurations, however we were unable to scale the number of actions and states higher for the last two configurations, as the synthesis run would not produce any results.

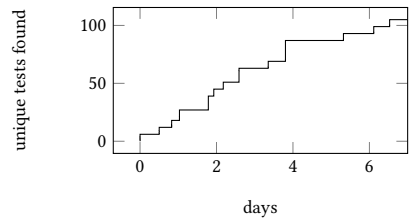
\begin{wrapfigure}{r}{6cm}
    \centering
    \scriptsize
    \begin{tikzpicture}
    \begin{axis} [xlabel={days},
       xmax=7,
       ylabel={unique tests found},
       width=5.5cm,height=3.5cm]
    \addplot [mark=none] table [
       x expr=\thisrow{time}/86400, 
       y=tests_found, 
       col sep=comma] {data/tests_generated_over_time.csv};
     \end{axis}
     \end{tikzpicture}
    \caption{The number of unique tests found over the 7 days. A total of 105 tests were found.} %(around 600K seconds) 

    \label{fig:tests-over-time}
\end{wrapfigure}
Our experiments are run with Alloy's symmetry breaking option, which heuristically aims to remove duplicate tests. However, we found that many duplicate tests were still generated. To address this, we wrote
a script to perform a textual diff after tests were converted into pseudo code; any tests found to be duplicates were discarded.

All synthesis runs (except the first) timed out, with a limit of 7 days. \autoref{fig:tests-over-time} shows the number of tests synthesized during this time for the 3-thread 4-instruction synthesis run. 

The rate of test synthesis appears to be slightly reducing after Day 4, which may suggest that the process is nearing completion, but we cannot know this with any certainty. It is also noteworthy that tests tend to be found in small bursts, with long periods of inactivity.

Timeouts and small bounds are typical in SAT-based synthesis work~\cite{mem-alloy}. Still, our synthesized tests are able to provide meaningful conformance test suites and provide illustrative examples (which we dub \textit{distinguishing tests}) for all-but-one of the progress models. Combined with testing heuristics, they are able to reveal interesting properties on real GPUs (\autoref{sec:heuristics}).

\subsection{Qualitative Analysis of Synthesized Tests} \label{sec:qualitative}

\begin{table}[]
    \centering
    \footnotesize
     \caption{The 5 tests that we manually identified in our synthesized test suite that correspond with classic idioms. We show the test (number of threads and instructions) as well as the LTS (number of actions and states). We also give the id, a sequentially increasing number given for each test in each synthesis run. }
    \begin{tabular}{l r r r r r}
    \toprule
    \textbf{Test Name} & \textbf{Threads} & \textbf{Instructions} & \textbf{ID} & \textbf{Actions} & \textbf{States} \\
    \midrule
        one-way producer-consumer (increasing id) & 2 & 2 & 5 & 3 & 3\\
        one-way producer-consumer (decreasing id) & 2 & 2 & 4 & 3 & 3\\
        bidirectional producer-consumer & 2 & 4 & 64 & 7 & 5\\
        simplified mutex & 2 & 3 & 14 & 7 & 6\\
        simplified dining philosophers & 2 & 2 & 0 & 8 & 8\\
        \bottomrule
        
    \end{tabular}
   
    \label{tab:qual-tests}
\end{table}

\begin{figure}
    \centering
        \scriptsize
        
        \begin{tabular}{l|l}
    \multicolumn{2}{c}{\textbf{one-way ProdCons (decreasing id)}} \\
    \toprule
    Thread 0 & Thread 1 \\
    \midrule
        \texttt{0: if (Mem[0] == 0)} &  \texttt{0: Mem[0] = 1} \\
         \texttt{\hphantom{xxxxx}goto 0} \\
        \bottomrule
    \end{tabular} \hspace{1cm}
                    \begin{tabular}{l|l}
    \multicolumn{2}{c}{\textbf{bidirectional ProdCons}} \\
    \toprule
    Thread 0 & Thread 1 \\
    \midrule
         \texttt{0: Mem[0] = 1} &  \texttt{0: if (Mem[0] == 0)} \\
        \texttt{1: if (Mem[0] == 1)} & \texttt{\hphantom{xxxxx}goto 0} \\
        \texttt{\hphantom{xxxxx}goto 1} & \texttt{1: Mem[0] = 0} \\
        \bottomrule
    \end{tabular}
    
        \vspace{.6cm}

    \begin{tabular}{l|l}
    \multicolumn{2}{c}{\textbf{simplified mutex}} \\
    \toprule
    Thread 0 & Thread 1 \\
    \midrule
         \texttt{0: if (Mem[0] == 1)}& \texttt{0: Mem[0] = 1}  \\
         \texttt{\hphantom{xxxxx}goto 0}& \texttt{1: Mem[0] = 0} \\
        \bottomrule
    \end{tabular} \hspace{1cm}
            \begin{tabular}{l|l}
    \multicolumn{2}{c}{\textbf{simplified dining philosophers}} \\
    \toprule
    Thread 0 & Thread 1 \\
    \midrule
        \texttt{0: if (Exch(Mem[0],0) == 1)} &  \texttt{0: if (Exch(Mem[0],1) == 0)} \\
         \texttt{\hphantom{xxxxx}goto 0} & \texttt{\hphantom{xxxxx}goto 0} \\
        \bottomrule
    \end{tabular} 
    \vspace{.6cm}

    \caption{Examples of synthesized tests that relate to idioms discussed in prior work on forward progress. The one-way producer-consumer test with increasing ids is shown in \autoref{fig:producer-consumer-early} and thus omitted from this figure. In all cases, memory locations are initialized to 0.}
    \label{fig:five-tests}
\end{figure}

Here we examine our synthesized test suite for \textit{qualitative} properties. That is, we gather interesting tests from prior works and search our synthesized tests for matches. We collect 5 tests, summarized in \autoref{tab:qual-tests} and shown explicitly in \autoref{fig:five-tests}. The one-way producer-consumer test with increasing id was shown earlier in \autoref{fig:producer-consumer-early}, and is thus ommited from \autoref{fig:five-tests}. Briefly, these five tests are: 

\begin{itemize}
    \item Three variants of Producer-Consumer (or ProdCons) in which one thread spins, waiting on a value from another thread. Prior work~\cite{Sorensen-Evrard-Donaldson-18} discussed that the thread ids associated with the producer and consumer thread determined whether the idiom was guaranteed to terminate under progress models like HSA and LOBE, as these models guarantee fair execution to threads with lower ids. We are able to synthesize ProdCons idioms where the producer has a larger id than the consumer, and vice versa. The third variant encodes bidirectional ProdCons, where the two threads wait on each other in sequence.

    \item A simplified dining philosophers test, in which each thread tries to store their id value to a memory location using an atomic exchange. If the thread sees the id of other thread was previously in the memory location, the it loops until it sees its own id. The two threads can livelock (taking turns storing their id), and thus this test requires a strongly fair scheduler for termination to be guaranteed.
    \item a simplified mutex test, in which T1 locks a location (by writing the value 1) and then unlocks the location (by writing the value 0). T0 will spin on a load instruction, waiting for the lock to be released (by observing 0). Of course, a robust mutex could not be written using the regular load/store sequences here, e.g.\ it would requre a read-modify-write instruction, however, the blocking behavior is similar to the blocking behavior seen in a mutex. 
\end{itemize}

This manual inspection of the tests, and relating them to common idioms in the literature, shows that our synthesis campaign was able to produce interesting tests.

\paragraph{Limitations and shortcomings}

\begin{wrapfigure}{T}{5.5cm}
    \centering
    \scriptsize

    \begin{tabular}{l|l}
    \multicolumn{2}{c}{\textbf{bidirectional ProdCons \#2}} \\
    \toprule
    Thread 0 & Thread 1 \\
    \midrule
         \texttt{0: if (Mem[0] == 0)} & \texttt{0: Mem[0] = 1}\\
        \texttt{\hphantom{xxxxx}goto 0} & \texttt{1: if (Mem[0] == 1)} \\
        \texttt{1: Mem[0] = 0} & \texttt{\hphantom{xxxxx}goto 1} \\
        \bottomrule
    \end{tabular} \hspace{1cm}
    \caption{Bidirectional ProdCons \#2 has similar blocking as bidirectional ProdCons \#1, i.e.\ both threads wait for each other, just in a different order. We consider this test redundant.\label{fig:qual-shortcomings}}
    
     \vspace{.5cm}  
        \includegraphics[height=4cm]{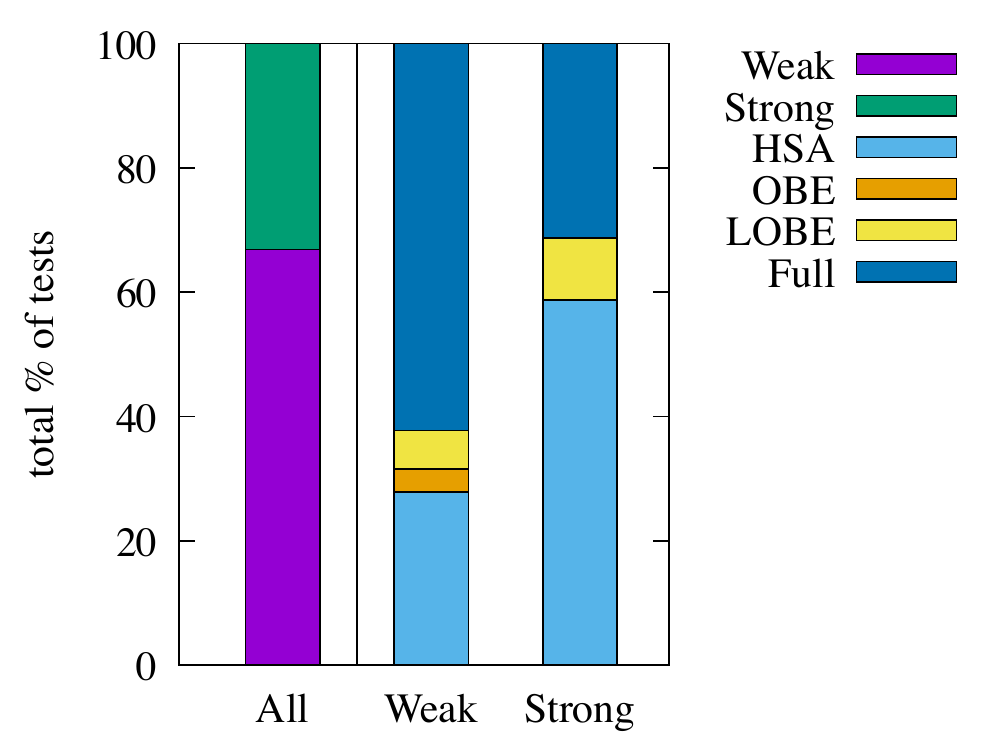}
    \caption{The distribution of tests across strong/weak schedulers \label{fig:strong-weak-scheduler-distribution}}
    
\end{wrapfigure}
Out of the set of tests we hoped to see in our synthesized test suite, there was one that we could not find: a traditionally implemented mutex, where each thread acquires the mutex properly (e.g.\ spinning on an atomic exchange) and then releases the mutex, by writing 0 to the mutux. This test was illustrated earlier in the paper in \autoref{fig:mutex}. We were unable to find this test although it should have been possible to synthesize under 2 threads and 4 instructions. Indeed if we work backwards from this test and supplement our Alloy model with constraints that match the mutex test, we were able to eventually synthesize it. Specifically, we added were precise constaints on the bounds  (8 states, 10 actions); a constraint that said each thread needed exactly 2 instructions; and a constraint that the first instruction (per-thread) must branch to itself, and the second instruction (per-thread) had no branch target.

Secondly, although we eliminate exact duplicates, our synthesis campaign can generate tests that might be redundant in the blocking synchronization they capture. For example, we synthesize two variants of the bidirectional ProdCons test (the first variant is shown in \autoref{fig:five-tests}, the second variant is shown in \autoref{fig:qual-shortcomings}). The only differences between the two tests is the order in which the two threads wait for each other. We cannot imagine a scheduler that would cause these two tests to behave differently, and thus we believe the tests are redundant. These two examples demonstrate that our technique does not have perfect \textit{precision} (i.e.\ some generated tests are not useful), nor does it have perfect \textit{recall} (i.e.\ some useful tests are not generated). However, we believe our results show that our technique is nonetheless \textit{useful} in practice.

\subsection{Running Tests on Formal Models}

We now analyze the behavior of our synthesized test suite under our executable formal models. To do this, we run each of the tests through each of our formal progress models implemented in the CADP model checker. For each test and progress model, we record whether the test passes or fails. The time to execute the CADP model checker across all these tests is short, as the tests are small and do not have large state spaces.

Using the results of the model checking, we can examine how our tests distribute over the formal schedulers. Our first analysis splits the tests into strong and weak tests: strong tests are guaranteed to terminate under a strongly fair scheduler, but may might not terminate under a weakly fair scheduler (i.e.\ due to a livelock). Weak tests are guaranteed to terminate under the weak scheduler. The percentage of our tests that are strong and weak are shown in the left most bar of \autoref{fig:strong-weak-scheduler-distribution}: \edit{about 65\% of our tests are weak, while the remaining are strong.}

In order to examine how our tests behave under semi-fair variants of the strong and weak fairness models, we first provide some definitions. A test $t$ for a progress model $m$ is a \emph{conformance test} if $t$ is guaranteed to terminate under $m$. A conformance test suite $S$ for model $m$ is useful because it can be run on an implementation $i$ of $m$ to check violations: any $t$ in $S$ that does not terminate would indicate that $i$ does not correctly implement $m$. Next we say that a test $t$ is \emph{distinguishing} for a model $m$ if $t$ is guaranteed to terminate under $m$, but $t$ is not guaranteed to terminate under all models that are strictly less fair than $m$. Thus, distinguishing tests are a subset of the conformance tests; and conformance tests are the union of distinguishing tests for $m$ and any progress model less fair than $m$. Distinguishing tests are useful in documenting the \emph{boundary} of a progress model. We note that we implicitly include the unfair scheduler in our analysis: by construction, none of our tests are guaranteed to terminate under the unfair model.

\begin{wraptable}{r}{6cm}
    \centering
    \scriptsize
     \caption{The number of distinguishing tests (D) and conformance tests (C) for each progress model. \edit{We split each model by its strong and weak variants.} \edit{The only model for which our synthesis approach did not produce any distinguishing tests is the strong variant of OBE.} }
    \begin{tabular}{l r r r r}
    \toprule
    & \multicolumn{2}{c}{\textbf{Weak Fairness}} & \multicolumn{2}{c}{\textbf{Strong Fairness}} \\
    \cmidrule(lr){2-3}\cmidrule(lr){4-5}
    \textbf{Model} & \textbf{D.\ Tests} & \textbf{C.\ Tests} & \textbf{D.\ Tests} & \textbf{C.\ Tests}\\
    \midrule
    HSA  & 90 & 90  & 94 & 94 \\
    OBE  & 12 & 12  & 0 & 0\\
    LOBE & 20 & 122 & 16 & 110\\
    full & 201 & 323 & 50 & 160\\
         \bottomrule
    \end{tabular}
    \label{tab:number-of-cd-tests}
\end{wraptable}
\autoref{fig:strong-weak-scheduler-distribution} shows the percentage of tests that are distinguishing for each of our semi-fair schedulers, broken down between the strong and weak tests. \autoref{tab:number-of-cd-tests} shows the concrete numbers,

as well as the number of conformance tests: e.g. 90 tests are guaranteed to terminate
 under the weak variant of HSA (henceforth called weak HSA), but not under the next less fair model in our hierarchy (i.e.\ the unfair model).Thus, these are distinguishing tests for the weak HSA. Because there are no distinguishing tests for a scheduler less fair than weak HSA, these 90 tests are also the conformance tests for weak HSA. As another example, the weak LOBE model has 20 distinguishing tests, i.e.\ tests that are guaranteed to terminate under weak LOBE, but not under weak HSA or weak OBE. The conformance tests for weak LOBE are the distinguishing tests for weak LOBE plus the conformance tests for weak HSA and weak OBE. 

We were able to gather distinguishing tests for all of our formal models, except for strong OBE. \edit{That is, while we have a formal specification of the strong OBE progress model, the synthesis did not provide any distinguishing tests for strong OBE (hence the 0 values in the strong OBE locations of \autoref{tab:number-of-cd-tests}). In future work, we hope to provide some distinguishing tests for this progress model, either by improving the efficiency of our test synthesis to automatically generate tests, or by constructing tests by hand and validating their behaviors using our executable models.}

\section{Empirical Testing} \label{sec:testing}

Now we transition into our empirical observations of running our tests on a variety of GPU hardware. Our main focus is testing the inter-workgroup scheduling behavior on GPUs, i.e.\ when threads are in different workgroups. This level of the GPU execution hierarchy is of particular importance
because although there exists very little official support or documentation for inter-workgroup synchronization, 
GPU programmers frequently write off-spec code that depends on forward progress properties.
This code is typically reported to work on the GPUs used for evaluation,
but there is no guarantee that it is portable across GPUs (or even that it will continue to work on the same GPU when the compiler or runtime environment changes).
One of our key results highlights this: through intensive testing we show that the Apple and ARM GPUs in our experimental campaign do \emph{not} support the LOBE progress model at the inter-workgroup level, refuting a hypothesis made in prior work~\cite{Sorensen-Evrard-Donaldson-18}.

\subsection{Heuristics for Provoking Interesting Behaviours} \label{sec:heuristics}

Our progress tests are small, consisting of only a small number of threads and instructions.
Prior work on GPU memory model testing using similarly small tests showed that interesting behaviors (relaxed memory behaviors) were extremely rare unless executed in a noisy environment, e.g. in the presence of additional threads that cause \emph{memory stress} e.g.\ by repeatedly accessing memory using irregular access patterns~\cite{Sorensen-Donaldson-16}.

Analogously, we have found it necessary to devise noisy environments as heuristics to help provoke interesting scheduler behaviours.
Prior work on inter-workgroup scheduling has shown that whether schedulers can cause non-terminating behaviours is dependent on the number of threads that are required to synchronize~\cite{owens-persistent,XF10,GPUBarrierOOPSLA2016}. For example, a barrier synchronization across all workgroups will hang due to a starvation cycle on a GPU if executed with too many workgroups. Inspired by this, we propose introducing \emph{scheduler stress} by having every kernel invocation execute many instances of a progress test simultaneously. Every instance of the test executes on its own set of disjoint workgroups and disjoint memory locations.
If the progress test specifies threads with ids $i$ and $j$, with $i < j$, our heuristics ensure that in each test instance, these threads will map to workgroups $w_i$ and $w_j$ with $w_i < w_j$ -- i.e.\ our stress heuristics preserve the relative id ordering in the progress test.

\begin{table}
    \footnotesize
    \centering
     \caption{The GPUs we consider in this study. We did not find more driver information for Apple devices apart from the iOS version. Due to an experimental error, we did not retrieve the driver version for the \mali.
     }
    \begin{tabular}{l l l l l l l l}
    \toprule
    \textbf{Name} & \textbf{Short Name} & \textbf{Framework} & \textbf{Vendor} & \textbf{Type} & \textbf{OS} & \textbf{Driver Version} \\
    \midrule
    GeForce 940m & \gfm & CUDA 9.1 & Nvidia & Discrete & Unbutu 18.04 & 440.100\\
    Quadro RTX 4000 & \rtx & CUDA 11.2 & Nvidia & Discrete & Ubuntu 20.04  & 460.32.03\\
    %\rowcolor{Gray} 
    A12 GPU & \aold & Metal & Apple & Mobile & iOS 14.4 & NA\\
    %\rowcolor{Gray} 
    A14 GPU & \anew & Metal & Apple & Mobile & iOS 14.4 & NA\\
    HD620 & \intel & Vulkan 1.2 & Intel & Integrated & Ubuntu 20.04 & Mesa 20.3.4\\
    Adreno 620 & \adreno & Vulkan 1.1 & Qualcomm & Mobile & Android 11 & RD1A.201105.003.C1\\
    Mali-G77 MP11 & \mali & Vulkan 1.1 & ARM & Mobile & Android 11& Missing\\
    Tegra X1 & \tegra & Vulkan 1.1 & Nvidia & Mobile & Android 9 & 1753219072\\
    \bottomrule
    \end{tabular}
   
    \label{tab:tested_gpus}
\end{table}

An $N$-thread progress test $t$ can be executed in one of three configurations:

\begin{enumerate}
    \item \textbf{Plain:} We launch a kernel with $N$ workgroups and map thread $i$ of the progress test $t$ to workgroup id $i$ ($0\leq i < N$). 
    
    \item \textbf{Round-robin:} We launch a kernel with $N\cdot M$ workgroups, for some $M > 1$, containing $M$ distinct instances $t_0, \dots, t_{M-1}$ of the progress test $t$.
    Test instance $t_m$ ($0\leq m < M$) operates on a distinct region of memory, indexed by $m$. Thread $i$ $(0\leq i < N$) of test $t_m$ is mapped to workgroup id $N\cdot m + i$. That is, threads are assigned to workgroups in a \emph{round-robin fashion}: threads for progress test $t_0$ run on the first $N$ workgroups; threads for progress test $t_1$ run on the next $N$ workgroups, etc.
    In general, a workgroup with id $w$ can determine its progress test id via $\left \lfloor{w/N}\right \rfloor$, and its thread id within the progress test via $w\;\mathrm{mod}\;N$.
    
    \item \textbf{Chunked:} As with round-robin, we launch $M$ distinct instances of the progress test via $N\cdot M$ workgroups, for some $M > 1$. This time thread $i$ ($0\leq i < N$) of test $t_m$ is mapped to workgroup id $M \cdot i + m$. That is, threads are organised into \emph{chunks} according to which thread of the progress test they correspond: the threads that correspond to thread 0 of each copy of the progress test run on the first $M$ workgroups; the threads that correspond to thread 1 run on the next $M$ workgroups, etc.
    In general, a workgroup with id $w$ can determine its progress test id $m$ via $w\;\mathrm{mod}\;M$ and its thread id within the progress test via $\left \lfloor{w/M}\right \rfloor$.
\end{enumerate}

Our chunked heuristic was designed specifically to tease out behaviors that distinguish LOBE from the fair model. For intuition, consider the producer-consumer with decreasing id shown in \autoref{fig:five-tests}. This test is \textit{not} guaranteed to terminate under LOBE as there are no guarantees that thread 1 will execute when thread 0 is spinning. When executing this test under the chunked heuristic, the kernel will have $M$ test instances and require $2*M$ workgroups. 

Workgroups 0 through $M$ will spin, waiting for the workgroups with ids $M$ to $2*M$ to be executed. The LOBE progress model hypothesizes a GPU scheduler that: (1)~is non-prememptive, i.e. restricted to executing at most $O$ workgroups concurrently; and (2)~schedules workgroups in linear id order. If $O < M$, then the GPU will initially schedule $O$ workgroups, with ids $0$ to $O$. These workgroups will all spin, waiting for starved workgroups that will not be eventually executed, because the scheduler is non-preemptive. 

Our round-robin heuristic does not have as much of a targeted intuition; it is simply a straightforward heuristic to implement. However, in the end, it was the key heuristic to observing our most surprising result: i.e.\ that Apple and ARM GPUs do not provide the LOBE progress model (\autoref{sec:lobe-conformance}).

\subsection{Methodology \label{sec:methodology}}

As summarized in \autoref{tab:tested_gpus}, we tested \numgpus{} GPUs, spanning \numvendors{} different GPU makers, including discrete, integrated and mobile GPU devices, under Linux, Android and iOS. Our tests cover the CUDA, Metal and Vulkan GPU programming frameworks.
We abandoned testing on devices from AMD because

repeated test time-outs
led to unrecoverable failures that sporadically required hard reboots.

For our CUDA GPUs, the \rtx was run on Ubuntu 20.04 using CUDA v.\ 11.2, the most recent. The \gfm was run on Unbutu 18.04 using CUDA v.\ 9.1. The \aold was run on an iPad Air 3 and the \anew was run on an iPhone 12, both running iOS 14.4.
The \intel was run on a Debian Linux with Mesa drivers v.\ 20.3.4.
The \adreno was run on a Pixel 5 and the \mali on a Galaxy S20-Exynos, both running Android 11.
The \tegra was run on an Nvidia Shield running Android 9.

We say that a test did not terminate if it times out. For CUDA, we use the Linux \texttt{timeout} command, using 20 seconds as our limit. For Vulkan we launch tests using the Amber testing framework~\cite{amber}, which allows a timeout to be set; this time we use a timeout of 5 seconds. For Metal, we could not find an iOS equivalent to the timeout feature, however, iOS implements a watchdog that kills non-terminating tests automatically. \edit{We could not find documentation specifying the timeout length, but visually, the display appears to freeze for 1-2 seconds.}

For each test, and each test variant (plain, round-robin, chunked), we run 20 iterations. The time to run our test suite varies substantially across devices.

We selected an iteration count that allowed all of our devices to finish running the test suite over night (e.g.\ within 12 hours), while still revealing interesting behaviors.

Kernels are launched with 1 thread per workgroup, thus all threads will be in different workgroups.
When the plain configuration is used, $N$ workgroups are launched, where $N$ is the number of threads participating in the test. When the round-robin or chunked heuristics are used,
$\left \lfloor{65,535/N}\right \rfloor$ instances of the test are launched so that up to 65,532 workgroups execute. The Vulkan specification requires devices to support at least this many workgroups~\cite[Table 53]{vulkan}, CUDA allows an even larger number of thread blocks (the CUDA equivalent of a workgroup)~\cite[App. I.1]{cuda-121}, and while we could not find any Metal documentation about allowed number of threadgroups (the Metal equivalent of a workgroup) we have found that this number of threadgroups is supported in practice.

\paragraph{Mapping \axb to GPU frameworks} Our mapping from the \axb language of \autoref{sec:programming-model} to the various GPU frameworks was not always straightforward. To target Vulkan, we used the OpenGL shading language (GLSL) frontend for Amber~\cite{amber}. The GLSL language does not provide support for \texttt{goto}; instead we implemented a program counter as a simple integer and emulated a \texttt{goto} environment using a \texttt{while} loop (one iteration per instruction) a \texttt{switch} statement (one case per instruction).  We kept this mapping for the rest of our languages.

\begin{wrapfigure}{R}{6cm}
    \centering
    \begin{lstlisting}[xleftmargin=.25in,basicstyle=\scriptsize\ttfamily,numbers=left]
int pc = 0;
while (pc != 2) {
  switch(pc) {
    case 0:
      if (atomicExchange(m,1)) {
        pc = 0;
      }
      else {
        pc += 1;
      }
      break;
    case 1:
      atomicExchange(m,0);
      pc += 1;
      break;
  }
}
\end{lstlisting}
    \caption{\edit{Snippet of the shader code (GLSL) of a thread of the mutex progress test (\autoref{fig:mutex})}}
    \label{fig:mutex-shader-code}
\end{wrapfigure}
\axb instructions map naturally to an atomic exchange instruction, which are available in all the backends we target. In the case where the exchange value is omitted (recall that this argument is optional), we use an atomic add of 0. This is an attempt to ensure all of our values make it to a point of coherence. Prior works on older GPUs showed that this was not always the case if only vanilla loads and stores were used~\cite{asplos-gpu-memory-model}. We provided memory model synchronization as-available, however many frameworks provide only limited support. For example, Metal is documented only to support relaxed consistency~\cite[p. 97]{metal}. GLSL does not provide atomic annotations that map to Vulkan's synchronization operations by default; although there is an extension to enable more synchronization annotations (\texttt{GL\_KHR\_memory\_scope\_semantics}), we opted not to use it to increase portability of our test suite.

The vast majority of our tests contain only a single memory location, thus coherence properties (supported even by relaxed atomic operations) should sequentialize memory accesses. For all of our surprising results (e.g.\ in \autoref{sec:lobe-conformance}), we have manually confirmed that memory consistency is not to blame.

\edit{To illustrate this compilation approach, \autoref{fig:mutex-shader-code} shows an snippet from the mutex progress test of \autoref{fig:mutex}, where both threads execute the same instructions.
In this snippet, notice that control flow is implemented as a while loop over a switch statement, where each case corresponds to an instruction. The program counter (\texttt{pc}) is updated once per instruction. In cases where there is branching (e.g.\ instruction 0), the \texttt{pc} is updated conditionally based on the result of the atomic exchange. In cases where there is no branching (e.g.\ instruction 1), the \texttt{pc} is simply incremented. The thread terminates (i.e.\ the while loop terminates) when the \texttt{pc} is incremented to one past the last instruction; in this example that value is 2, as there are 2 instructions.}

\subsection{Empirical Testing Results for Inter-workgroup Schedulers} \label{sec:inter-workgroup-testing}

\begin{figure}

    \centering
\includegraphics[width=\linewidth]{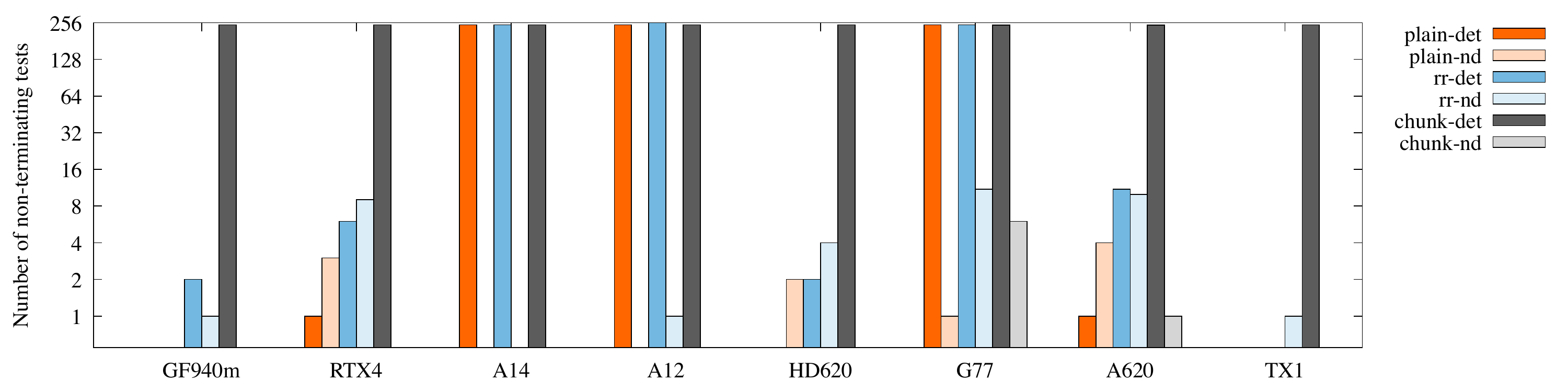}
\caption{Summary of non-terminating test observations across our GPU devices. 
}
    \label{fig:inter-workgroup-violations}
\end{figure}

\subsubsection{Inter-workgroup Schedulers are Empirically Less Fair Than CPU Schedulers}

It has been well-documented that GPU inter-workgroup schedulers are less fair than CPU schedulers, e.g.\ see~\cite{GPUBarrierOOPSLA2016, XF10, owens-persistent}. Thus, a good initial experiment is to see if we observe non-terminating tests in our test suite when executed on GPU platforms, and if all of our tests terminate when executed on a CPU platform. To do this, we wrote a C++ backend for our test suite using atomic operations and std threads. 

We executed the three variants (plain, round-robin and chunked) of our 483 tests across all GPUs in \autoref{tab:tested_gpus} and a CPU (an 8-core i7-9700K). For the CPU, we ran two configurations: fully-subscribed with as many threads as cores (i.e.\ 8) and over-subscribed with 100$\times$ as many threads as cores (i.e.\ 800). The over-subscribed configuration is intended to assess the fairness provided by CPU schedulers when cores are over-occupied. On the other hand, GPUs have been conjectured to guarantee fair scheduling only up to the number of threads that occupy their hardware resources~\cite{Sorensen-Evrard-Donaldson-18}. We set the CPU timeout to the same as the CUDA tests: 20 seconds.

For all variants of all tests, the tests executed on the CPU terminated within the time limit. These results are in-line with CPU programming folklore and the recommended progress model provided by the C++ specification~\cite[Sec. 4.7.2]{C++-spec}. Specifically, the specification states that implementations are encouraged (but not required) to provide std threads with ``concurrent forward progress,'' which we read to be equivalent to weak fairness. While the C++ specification says nothing about strong fairness, the CPU execution environment provides enough timing variation (e.g.\ due to background processes) that all actions are eventually taken. 

The GPU test results were more varied: \autoref{fig:inter-workgroup-violations} summarizes our results, shown
per-GPU (on the x-axis), per variant, and whether or not the non-termination observation was deterministic (i.e. all 20 iterations did not terminate). The results are shown on a log axis as there were some variants with a high number of non-terminating instances, while others variants exhibited non-terminating behavior on far fewer tests. Each GPU has slightly different behavior across our test suite: The Apple GPUs and \mali deterministically exhibit non-termination across roughly the same number of tests. The older Nvidia GPUs (\gfm and \tegra) had very few non-terminating tests on the plain variants. The \rtx, \adreno, and \intel  have very similar profiles, despite all 3 being from different vendors. We will examine these results in more detail in the next two subsections.

\subsubsection{Do Inter-workgroup Schedulers Conform to the LOBE Progress Model?} \label{sec:lobe-conformance}
We now check whether our GPUs conform to the LOBE progress model, as has been hypothesized~\cite{Sorensen-Evrard-Donaldson-18}, i.e.\ we check whether LOBE conformance tests terminate across our GPUs. We consider the conformance tests for \emph{weak} LOBE; we do not consider strong progress here because it seems unlikely that it will be included in a language specification. From \autoref{tab:number-of-cd-tests}, we see there are 122 tests conformance tests for weak LOBE, for which we have 3 variants each: plain, chunked, and round-robin. This is a combination of 366 tests, each executed for 20 iterations.

Our analysis shows that all GPUs pass all weak LOBE tests, with the exception of the \aold and \mali. There are 11 weak LOBE conformance tests for which both of these GPUs exhibit non-termination. Interestingly it is the same set of 11 tests across the three GPUs; however, the \aold is more deterministic, exhibiting non-termination deterministically on 10 out of the 11 tests. The \mali does not deterministically exhibit non-termination on any of these tests, but it does occur on over half of the runs on all 11 tests. We did not have intuitions that round-robin would be an effective heuristic, but all the violating 11 tests have the round-robin heuristic enabled. The reason for this is that LOBE violations require that threads with a higher id have a chance to execute before threads with a lower id. One scenario that violates LOBE is this: (1)~a thread with a high id executes a step; (2)~this promises eventual execution for all threads with lower ids; (3)~one of the threads with the lower id isn't fairly scheduled. Step (1) requires a workgroup with a high id to be initially scheduled. With the chunked heuristic this may not happen, as GPUs are likely schedule workgroups in some kind of ascending order, but round-robin maps all threads of a progress test to workgroups with adjacent ids, giving workgroups with a higher id the chance to initialize this scenario.  

Apart from our large experimental campaign, we have also observed rare LOBE violations when experimenting informally with the \anew GPU: this occurs when running only the LOBE conformance tests. We could not reproduce these results when running the entire testing suite. We hypothesize that the larger testing suite causes the GPU to lock up more frequently, putting the GPU in a constant state of recovery, which inhibits some of the behavior we might observe. Furthermore, our contacts at Apple provided an additional test that is able to show a violation of OBE on the A12 and A14. This test consists of threads repeatedly acquiring and releasing a mutex using a custom spin-lock (implemented using an atomic compare-and-swap). In future work, we hope to examine this test in depth. Specifically, we aim to decouple the idiomatic blocking behavior from the system stress caused by the mutex contention. At that point, we could apply the stress as a testing heuristic across all of our tests to more robustly test many more progress behaviors.

Thus, the hypothesis of prior work~\cite{Sorensen-Evrard-Donaldson-18} does not hold for these devices, which span two frameworks: Metal and Vulkan, hence these frameworks will both need to consider how to address these violations when specifying forward progress properties: it may be that LOBE could be provided as an optional extension, as it appears to be empirically supported on many devices. We have informed representatives from both Apple and ARM about these behaviors. 

We did not observe any violations of the weak variant of HSA or OBE on any devices, which suggests that it may be possible for a specification to minimally require one of the less fair models (e.g.\ HSA or OBE), with an optional extension for schedulers that are more fair (e.g.\ LOBE).

\subsubsection{Uniform Behaviors Under the Chunked Heuristic} \label{sec:distinguishing-lobe} 

We now discuss the clear uniform result across our GPUs: roughly 256 tests deterministically fail using the chunked heuristic. We hypothesize that this is because GPU workgroup schedulers are non-preemptive and assign workgroups to compute units in linear order. As described in \autoref{sec:heuristics}, this is exactly the type of scheduler that the chunked heuristic was designed to provoke non-terminating behaviors from. Because LOBE was developed to describe such a scheduler, we imagine that these empirically failing tests correspond to tests that distinguish fair schedulers (i.e.\ fail under LOBE but pass under fair). From \autoref{tab:number-of-cd-tests} there are 251 tests that distinguish the fair from LOBE (combining strong and weak variants). Our analysis shows that the chunked variants of these tests have non-terminating behavior across all GPUs for the same subset of 246 of these tests.

Further examination of the 5 tests that distinguish full fairness from LOBE, but that deterministically terminate across all our GPUs, reveals an interesting pattern: these tests all have 3 threads and they require threads 0 and 1 to execute in parallel in order to get in the endless starvation cycle, which consists of thread 2 being starved in favor of thread 0. However, our chunked heuristic is designed so that none of the testing threads will execute in parallel on non-preemptive systems; thus we do not observe non-termination for these tests. A hybrid or randomized mapping strategy, e.g.\ that maps the first two threads in a round-robin style, and the third thread in a chunked style, might be capable of revealing non-terminating behavior for these tests.

\subsubsection{Inter-workgroup Strong Fairness} The remaining failing tests are strong tests, including for less fair schedulers, e.g.\ strong HSA and strong OBE. While programming languages do not (and perhaps should not) provide strong fairness guarantees, it is interesting that the strong tests terminate reliably when executed on a CPU, even when heavily oversubscribed. Our results show that GPUs are more prone to livelock, likely due to
the absence of background processes or a complex OS scheduler to introduce 
timing jitters that break threads out of livelock.
%Thus, GPU workgroups seem more likely to suffer from livelock than threads in a traditional concurrent CPU environment.

\subsection{Limitations and Additional Results} \label{sec:additional-results}

The main focus of this work is inter-workgroup progress. As such, we consider progress in a \textit{Multiple Instruction, Multiple Data} (MIMD) context. This may seem counter-intuitive, as GPUs are programmed in a Single Program, Multiple Data (or SPMD) manner. However, a MIMD programming model can be implemented in a SPMD model by using a set of conditionals that branch on thread ids; thus we regard the difference, for the purpose of this work, as superficial. Furthermore, to the best of our knowledge, GPU threads that are in different subgroups do execute in a MIMD fashion, i.e.\ containing independent program counters with execution that is not lock-step synchronous. 

The techniques in this work may not be appropriate for all levels of the GPU execution hierarchy. For example, threads in the same subgroup might execute in a synchronous, lock-step manner, also known as \textit{Single Instruction, Multiple Thread} (SIMT). Threads that don't execute the same instruction, i.e.\ they have control flow divergence, are sequentialized. Thus, our tests and models, which are control flow agnostic, are not suitable for describing interactions at this level.

To the best of our knowledge, CUDA is the only framework that provides documentation for progress, and its guarantees are limited to situations out of our main focus: intra-subgroup, and a special \textit{cooperative groups} API. 
Nonetheless, because it is straightforward, we did run our tests in these configurations and we now provide a high-level overview of the results:

\paragraph{CUDA cooperative kernels} CUDA's cooperative kernel launch~\cite[App. C.7]{cuda-121} claims to allow workgroups to ``cooperate and synchronize as they execute'', which we believe provides sufficient progress across workgroups to support common synchronization idioms such as mutexes and barriers. We validated this claim by executing our test suite on the \rtx (our only GPU that supports this feature), observing termination for all weak tests. However, we did not observe termination for all strong tests. Thus, the cooperative kernel launch remains empirically weaker than our observations on a traditional CPU, due to livelocks.

\paragraph{Independent warp scheduler} Nvidia GPUs document an independent warp scheduler (IWS) since the Volta architecture~\cite{VoltaWhitepaper}. Similar to the cooperative kernel launch, we hypothesize that this scheduler provides a weak progress model. To test this hypothesis, we created a suite of our \textbf{plain} variation tests that target intra-warp threads on the \rtx GPU, which supports the independent warp scheduler. We empirically validate our hypothesis by observing termination for all weak tests. However,  we did not observe termination for all strong tests. 

In a final experiment, we ran the intra-warp tests on the same device (\rtx) under the SIMT warp scheduler (toggled with a compiler flag). These test results show failures on some weak tests (as expected due to SIMT execution), however the sequentialized execution is able to break some livelock cycles found in strong tests. We found several tests that reliably pass under the SIMT warp scheduler, but not the IWS, and vice versa. Thus, empirically, the two schedulers are incomparable.

\section{Related work}\label{sec:related}

\paragraph{GPU progress models} Several prior works have suggested progress models for GPUs. Early work discussed a \textit{maximal launch}, in which CUDA kernels would be launched as many workgroups as the device could concurrently execute~\cite{owens-persistent}.

As discussed throughout the paper, \citet{Sorensen-Evrard-Donaldson-18} proposed a family of \textit{semi-fair} schedulers to describe inter-workgroup schedulers and hypothesized that current GPUs implement the linear occupancy-bound execution model; a hypothesis we refute with empirical evidence in \autoref{sec:testing}.
A different approach discusses architectural and compiler techniques that could make it easier for GPUs to provide fair progress models for workgroups~\cite{Dutu-Sinclair-Beckmann-Wood-Chow-20}. 
Although it is not the main focus of this work, other works have formalized SIMT-style semantics to intra-subgroup threads~\cite{DBLP:conf/esop/HabermaierK12,gpuverify,DBLP:journals/toplas/BettsCDKQTW15,gklee,DBLP:conf/esop/CollingbourneDKQ13}, which provide a predicated execution model. To avoid specifying less fair models for intra-subgroup threads, ~\citet{MIMDSyncOnSIMD} propose architecture and compiler techniques to provide fair models to these threads.

\paragraph{Demystifying GPU behaviors through testing}  There is precedence for using large testing campaigns to demystify system behaviors, especially GPU systems. Most closely related is~\citet{asplos-gpu-memory-model}, who performed a large testing campaign to understand the semantics of Nvidia GPU memory models. In a different dimension, there is a line of work, starting with~\citet{demystifying-gpu}, that aims to discover performance-critical architectural details through microbenchmarks.

\paragraph{Conformance test generation for GPU programming frameworks} The GLFuzz project~\cite{glfuzz} generates compiler tests for a variety of GPU frameworks, some of which are now incorporated into Khronos Group conformance tests~\cite{PuttingRandomizedCompilerTestingIntoProduction}. However, GLFuzz does not consider fine-grained synchronization behaviors, as we do in this work. Other works have generated tests for memory model conformance~\cite{mem-alloy, lustig+17}, including for Nvidia GPUs and OpenCL. Along these lines, ~\cite{amd-coherence-testing} describes a random test generator for coherence on AMD GPUs.

\paragraph{Model checking concurrent program termination.}
There is vast literature on model checking for concurrent program termination. \citet{MateescuSerwe10} provides an extensive study of termination, analyzing the performance of various shared memory mutual exclusion protocols, taking both weak and strong fairness into account.
In another vein, \citet{lahav2020making} discuss liveness properties in the context of relaxed memory models, exploring how to specify that memory values eventually propagate through the memory system and become visible to other threads.

\section{Conclusion}\label{sec:conclusion}

This work presents techniques, tools, and experimental results to aid programming language designers in thinking about progress models. Our individual contributions allow developers to specify progress models in executable semantics, synthesize test suites, and explore behaviors on existing GPU implementations. When combined, these contributions allow synthesized tests to be partitioned into conformance test suites, which, in turn, can be used to analyze empirical results to determine if certain GPUs experimentally conform to a given progress model. Our results highlight the power of this synergy by showing that Apple and ARM GPUs do not implement a progress model hypothesized in prior work. We are in discussion with industrial representatives about how to incorporate progress models into official specifications and how the corresponding test suites might be incorporated into conformance tests. 
Our hope is that this work inspires other areas of specification design to embrace formal methods, both for executable semantics and for test-case synthesis, to explore design decisions and obtain rigorous conformance test suites.

\section*{Acknowledgments}
We thank the anonymous reviewers for their feedback, which greatly improved the clarity of the paper. We give special thanks to Alan Baker (Google) who provided detailed support for the Amber framework and gave many valuable comments on a draft of the paper. We thank the Khronos SPIR Memory Model TSG, especially Rob Simpson (Qualcomm), David Neto (Google), Jeff Bolz (Nvidia), Nicolai Hähnle (AMD), Graeme Leese (Broadcom), Brian Sumner (AMD), Tobias Hector (AMD), and Mariusz Merecki (Intel) for their support and feedback on this work over several years. 

We also thank the Inria Convecs team that maintains CADP for their support, especially Radu Mateescu for his feedback on our MCL formulas. This work was partially supported by the EPSRC via the IRIS Programme Grant (EP/R006865/1) and the HiPEDS Doctoral Training Centre (EP/L016796/1).

%\newpage

\bibliography{references}
%\end{document}

\ifx\includeappendix\undefined
\end{document}
\fi

\newpage

\section{Appendix}

\subsection{Proofs of non-termination detection conditions}

\subsubsection{Weak Fairness}

\begin{lemma}
In the case of weak schedulers, a given test case has the possibility of deadlock if and only if it contains a cycle that has executing steps by the all threads in the UNIQUE set of fairly scheduled threads $S$ shared by every step in the cycle.
\end{lemma}

\begin{proof}
The first thing we will see is that the definition is well-defined, and all actions in a cycle must have the same set of fairly scheduled threads $S$. To see this, note that for all the schedulers in this work, the set of fairly scheduled threads only removes a thread when it terminates. Since a terminated thread will not execute again, it follows the set cannot remove and add a thread. Thus, the set of fairly scheduled threads $S$ in a cycle must be the same.

We can then note that, if the output graph for a test under a given scheduler contains a cycle with all threads in $S$ at the time of the cycle, then we are in a deadlock situation. This follows from the fact that only those threads are guaranteed to execute and, thus, no other thread that could break execution from that cycle will run.

It follows that we must only see that the presence of a deadlock implies one of these cycles. We will show this by contrapositive. Let us assume the lack of problematic cycles, we will then show we can reach termination. To do this, note once again that any cycle must have a fixed $S$. Let us call the set of nodes (and edges between them) with this fixed set $S$, $G$. We will show now that if we do not have a problematic cycle, then we must move on to a different part of the graph with another $S'$ where a thread was added to $S$ or finished execution (and thus removed from $S$).

If we have no cycles, then this is clearly true because we have finite states. Else, let us take a cycle $C$ that includes as many threads in $S$ as possible. Note because $C$ is not problematic, there must be a $t \in S$ that is not in the cycle and, eventually, $t$ must execute and lead us to break out of that cycle. Two things are then possible. Either we end up in another cycle $C'$ with the same $S$, or there are no more cycles with this $S$ from here and we eventually add or remove a thread from $S'$ (as we wanted). If the former happens, note that we cannot reach any of the nodes in the earlier cycle (as else $C$ would not be maximal as we can add $t$ into the cycle). But, note then $C'$ is a cycle in a subgraph $G'$ with strictly less nodes than $G$. After a finite number of steps (since $G$ is finite), it must then be impossible to find a graph in $G'$ as it becomes empty. It follows then $S$ must change, implying we add a node to it or a node terminates (and is removed).

Thus, we must eventually add fairly scheduled threads or terminate them. Since the number of threads is finite and each is added and removed at most once to $S$, this must mean eventually all threads will need to terminate, giving us that we must reach program termination (as we wanted).
\end{proof}

\subsubsection{Strong Fairness}

\begin{lemma}
In the case of strong schedulers, a given test case is free of deadlock if and only if for every node there exists a possible path to the termination of the fairly scheduled threads (the set is empty) such that the path contains actions taken only by fairly scheduled threads.
\end{lemma}

Note we make the distinction that we reach a state where the set of fairly scheduled threads is empty (instead of reaching exit) as in OBE we only fairly scheduled threads when they execute and, in the exact moment when we finish executing the current set, the set is empty, even though it will add a thread to the set in the next action.

\begin{proof}
For the first direction, we note that if the path does not exist then we have a possible deadlock situation. To see this, note that means we have a node $x$ such that, if we explore the graph taking only actions by fairly scheduled threads, we never reach the termination of all these threads (and any threads that may get added).Because our set of fairly scheduled threads is empty only when all the threads that were added to the set terminated, if the search did not reach this state, it must mean that our search terminated because all possible interleaving ended in a cycle (as every node has to have one action from every unfinished fairly scheduled thread). It follows that no matter the scheduling of the fairly scheduled threads we reach a cycle. Furthermore, no other threads are guaranteed to intervene so this deadlock could happen.

Then, for the second direction, we must now see that if this path exists for every node we can reach termination of the current threads. We do so by contradiction. Let us assume that we still have a cycle. Then, let us take the set of all nodes/actions which we visit/take infinitely often in the cycle. We know there is a path to exit from any of these nodes using only fairly scheduled threads, take one such path from a node. Note that this path has to have a first action that is not taken in the cycle eventually (else we would have reached this exit). Because this is the first new action in the path, we know we must reach the state from which it can happen infinitely often. Thus, this action must happen infinitely often, which is a contradiction as then our set of infinitely often visited nodes/actions would be larger. It follows that we must eventually reach the termination of all these nodes if there is such an exit (the only node with no outgoing edges).
\end{proof}

\subsection{Forward Progress Execution Models}
\label{sec:lnt-progress-models}

We present the LNT specification of the HSA, HSA+OBE, LOBE, unfair and fair progress models.

%For completeness, we include the LNT code for each of our specified progress models. Note the strong variant determination is not given by the LNT code, but rather by MCL formulas. Once again, as before, we omit the termination conditions for brevity.

\begin{figure}[H]
\begin{lstlisting}[language=LNT,xleftmargin=.25in,basicstyle=\scriptsize\ttfamily]
process HSA [Step: ExecutionStep, Terminate: Natural] is
  var
    tid: Nat,      -- thread ID
    axb: AxbInst,  -- AXB instruction
    F:   NatSet,   -- set of threads guaranteed fair execution
    smallest: Nat, -- smallest active thread
    done: NatSet   -- set of terminated threads
  in
    smallest := 0; -- Initially, the smallest non-terminated thread id is 0.
    F := {smallest};
    done := {};
    
    loop
      select -- non-deterministic choice operator
        Step(?tid, ?axb, F)  -- some thread executes a step
      []
        Terminate(?tid);     -- thread tid has terminated its own program
        done := insert(tid, done); -- remember this thread as terminated
        
        while member(smallest, done) loop  -- get next smallest non-terminated thread id
            smallest := smallest + 1
        end loop;
        F := {smallest} -- F contains the smallest, non-terminated thread id

      end select
    end loop
  end var
end process
\end{lstlisting}
\caption{Specification of the HSA progress model in the LNT formal language.}
\label{fig:lnt-hsa}
\end{figure}

\begin{figure}[H]
\begin{lstlisting}[language=LNT,xleftmargin=.25in,basicstyle=\scriptsize\ttfamily]
process HSA_OBE [Step: ExecutionStep, Terminate: Natural] is
  var
    tid: Nat,      -- thread ID
    axb: AxbInst,  -- AXB instruction
    F:   NatSet,   -- set of threads guaranteed fair execution
    smallest: Nat, -- smallest active thread
    done: NatSet   -- set of terminated threads
  in
    smallest := 0; -- Initially, the smallest non-terminated thread id is 0.
    F := {smallest};
    done := {};

    loop
      select -- non-deterministic choice operator
        Step(?tid, ?axb, F); -- thread tid executes a step
        F := insert(tid, F)  -- thread tid is now granted fair execution guarantee
        
      []
        Terminate(?tid);     -- thread tid has terminated its own program
        F := remove(tid, F); -- remove thread tid (if in F) from F
        done := insert(tid, done) -- mark thread tid as done
        
        while member(smallest, done) loop -- get next smallest non-terminated thread id
            smallest := smallest + 1
        end loop;
        F := insert(smallest, F) -- make sure this thread id is fairly executed
        
      end select
    end loop
  end var
end process
\end{lstlisting}
\caption{Specification of the HSA+OBE (the intersection of both) progress model in the LNT formal language.}
\label{fig:lnt-hsa-obe}
\end{figure}

\begin{figure}[H]
\begin{lstlisting}[language=LNT,xleftmargin=.25in,basicstyle=\scriptsize\ttfamily]
process LOBE [Step: ExecutionStep, Terminate: Natural] is
  var
    tid: Nat,      -- thread ID
    axb: AxbInst,  -- AXB instruction
    F:   NatSet,   -- set of threads guaranteed fair execution
    done: NatSet,  -- set of terminated threads
    t: Nat         -- iteration variable
  in
    F := {};       -- At the beginning, no thread is guaranteed fair execution
    done := {};

    loop
      select -- non-deterministic choice operator
        Step(?tid, ?axb, F); -- thread tid executes a step
        -- All non-terminated threads with an id lower or equal to tid are
        -- guaranteed fair execution
        for t := 0 while t <= tid by t := t + 1 loop
          if not(member(t, done)) then
            F := insert(t, F)
          end if
        end loop
      []
        Terminate(?tid);           -- thread tid has terminated its own program
        F := remove(tid, F);       -- remove thread tid (if in F) from F
        done := insert(tid, done)  -- mark thread tid as terminated
      end select
    end loop
  end var
end process
\end{lstlisting}
\caption{Specification of the LOBE progress model in the LNT formal language.}
\label{fig:lnt-lobe}
\end{figure}

\begin{figure}[H]
\begin{lstlisting}[language=LNT,xleftmargin=.25in,basicstyle=\scriptsize\ttfamily]
process UNFAIR [Step: ExecutionStep, Terminate: Natural] is
  var
    tid: Nat,      -- thread ID
    axb: AxbInst,  -- AXB instruction
    F:   NatSet    -- set of threads guaranteed fair execution
  in
    F := {};       -- No thread is ever guaranteed fair execution
    loop
      select -- non-deterministic choice operator
        Step(?tid, ?axb, F)
      []
        Terminate(?tid)
      end select
    end loop
  end var
end process
\end{lstlisting}
\caption{Specification of the unfair progress model in the LNT formal language.}
\label{fig:lnt-unfair}
\end{figure}

\begin{figure}[H]
\begin{lstlisting}[language=LNT,xleftmargin=.25in,basicstyle=\scriptsize\ttfamily]
process FAIR [Step: ExecutionStep, Terminate: Natural] (max_tid: Nat) is
  var
    tid: Nat,      -- thread ID
    axb: AxbInst,  -- AXB instruction
    F:   NatSet    -- set of threads guaranteed fair execution
  in
    -- Each thread is guaranteed fair execution, until it terminates
    F := {};
    for tid := 0 while tid < max_tid by tid := tid+1 loop
        F := insert(tid, F)
    end loop;
    
    loop
      select -- non-deterministic choice operator
        Step(?tid, ?axb, F);  -- thread tid makes a step
      []
        Terminate(?tid);      -- thread tid has terminated its own program
        F := remove(tid, F)   -- remove thread tid (if in F) from F
      end select
    end loop
  end var
end process
\end{lstlisting}
\caption{Specification of the fair progress model in the LNT formal language. This one is a bit special as it needs to know the total amount of threads, in order to initialize $F$ with all thread ids at the beginning.}
\label{fig:lnt-fair}
\end{figure}

\ifdefined\includeappendix
\end{document}
\fi